\documentclass[final,5p,times,twocolumn,authoryear]{elsarticle}

\usepackage{amsmath, amssymb, amsfonts, amsthm, fouriernc, mathtools}
\usepackage{gensymb} 
\usepackage{float}
\graphicspath{{figures/}}
\usepackage[]{units}
\usepackage{caption}
\usepackage{subcaption}
\usepackage[usenames, dvipsnames]{color} 
\usepackage{multirow}
\usepackage{eurosym}
\usepackage{tabularx}
\usepackage{pdflscape}
\usepackage{xcolor}
\usepackage{makecell}
\usepackage{pdfpages} 
\usepackage{changepage} 
\usepackage{xr}
\usepackage{xfrac} 
\journal{Energy Policy}

\makeatletter
\def\ps@pprintTitle{%
 \let\@oddhead\@empty
 \let\@evenhead\@empty
 \def\@oddfoot{\centerline{\thepage}}%
 \let\@evenfoot\@oddfoot}
\makeatother

\begin{document}

\begin{frontmatter}


\title{Impact of climate change on the cost-optimal mix of decentralised heat pump and gas boiler technologies in Europe}

\author[rvt,rvt2]{S. ~Kozarcanin\corref{cor1}}
\ead{sko@eng.au.dk}

\author[rvt3]{R. ~Hanna}
\ead{r.hanna@imperial.ac.uk}

\author[rvt2,rvt3]{I. ~Staffell}
\ead{i.staffell@imperial.ac.uk}

\author[rvt3]{R. ~Gross}
\ead{robert.gross@imperial.ac.uk}

\author[rvt]{G. B. ~Andresen}
\ead{gba@eng.au.dk}

\cortext[cor1]{Corresponding author}

\address[rvt]{Department of Engineering, Aarhus University, Inge Lehmanns Gade 10, 8000 Aarhus, Denmark}

\address[rvt2]{Imperial College Centre for Environmental Policy, 16--18 Prince's Gardens, South Kensington, London SW7 1NE, UK}

\address[rvt3]{Imperial College Centre for Energy Policy and Technology, 16--18 Prince's Gardens, South Kensington, London SW7 1NE, UK}

\address{}

\begin{abstract}
Residential demands for space heating and hot water account for 31\% of the total European energy demand. Space heating is highly dependent on ambient conditions and susceptible to climate change. We adopt a techno-economic standpoint and assess the impact of climate change on decentralised heating demand and the cost-optimal mix of heat pump and gas boiler technologies. Temperature data with high spatial resolution from nine climate models implementing three Representative Concentration Pathways from IPCC are used to estimate climate induced changes in the European demand side for heating. The demand side is modelled by the proxy of heating-degree days. The supply side is modelled by using a screening curve approach to the economics of heat generation. We find that space heating demand decreases by about 16\%, 24\% and 42\% in low, intermediate and extreme global warming scenarios. When considering historic weather data, we find a heterogeneous mix of technologies are cost-optimal, depending on the heating load factor (number of full-load hours per year). Increasing ambient temperatures toward the end-century improve the economic performance of heat pumps in all concentration pathways. Cost optimal technologies broadly correspond to heat markets and policies in Europe, with some exceptions. \end{abstract}


\end{frontmatter}

\newpage
\section{Introduction}

\noindent Energy consumption for space heating is by far the most important end-use in the European (EU28) residential heating sector with an estimated share of 52\% in 2015 \citep{HRE4}. Space heating is strongly temperature dependent and mostly consumed during cold winter seasons \citep{kozarcanin2019estimating}. Heating systems are therefore designed to meet peak demand during cold winter periods, but for long-term design decisions, it is necessary to focus on long-term changes in the climate. Depending on the degree of climate change in the future, it is believed that the peak demand for space heating might change significantly. Given these points, the principal aim of this study is to analyse the 21st Century climate change impact on the selection of cost-optimal, decentralised heating technologies for different locations in Europe. We define decentralised heating as all heating systems installed on a per-building basis. This means that we do not focus on large-scale centralised heating systems such as combined heat and power plants or other district heating facilities. This paper draws upon climate affected temperature data from the newest simulations carried out in the framework of the CMIP (Coupled Model Inter-comparison Project) Phase 5 \citep{taylor2012overview, flato2013evaluation} and the EURO-CORDEX project \citep{kotlarski2014regional, jacob2014euro}. We present nine climate models from a combination of 4 regional climate models, RCM, downscaling 5 global climate models, GCM, under the forcing of the latest generation of climate projections provided by the Intergovernmental Panel on Climate Change, IPCC \citep{moss2010next}. For this study, we use the best available resolutions which is 3hr in time and 0.11$\degree$ $\times$ 0.11$\degree$ in space for Europe. A full description of the climate data is provided in the Supplementary Information Section 4 (SI 1.4). \\

\noindent A fundamental impact on the selection of heating technologies, that to the best of the authors knowledge has not been studied in detail, is the impact of local climates on the cost-optimal design of decentralised heating systems. Throughout this article, we use \emph{system design} to refer to the cost-optimal selection of decentralised heat generating technologies. Spatial variations in the ambient temperatures fluctuate heterogeneously from the oceanic to the mainland climates of West and East Europe, respectively, and from the cold northern to the warmer southern climates. Climate change is furthermore expected to introduce long term and heterogeneous temperature anomalies across Europe. Whereas hot water demand is relatively constant throughout the year and between years \citep{staffell2015domestic}, the energy consumed for space heating will therefore fluctuate more wildly and be subject to long-term trends that are currently not well understood. \\

\noindent A secondary aim of this paper is to evaluate the fit between cost-optimal technologies for decentralised heating and heat policies in Europe. Actual deployment of heating technologies in different countries may not necessarily reflect which technologies are most cost-optimal in a given location. The purpose of this policy assessment is to identify where policy intervention might be required to achieve lower cost outcomes, while contributing to overall efforts to decarbonise heating. \\

\begin{figure*}
    \centering
        \includegraphics[width=1\textwidth]{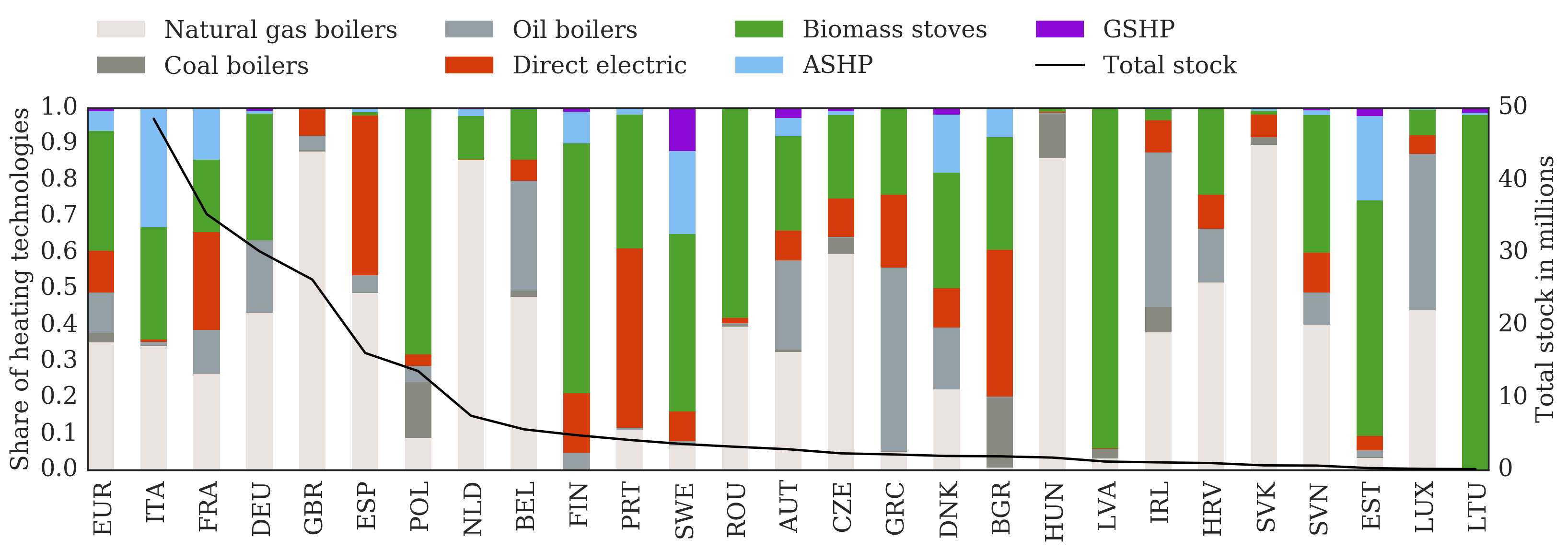}
    \caption{Shares of the installed stock of heating technologies across the European countries in 2012. The x-axis lists the countries, referred to by their three-letter ISO codes. Data from \cite{fleiter2016mapping}.} 
    \label{Fig: tech_share}
\end{figure*}

\noindent Fig. \ref{Fig: tech_share} illustrates the current technology shares that are responsible for delivering the decentralised heat for a majority of the European countries. The European average bar shows that fossil fueled boilers dominate the heat generation, followed by biomass fueled technologies. Heat pumps are relatively new technologies compared to boilers and consequently hold a minor share of the total installed technology stock. On the other hand, heat pumps are gaining more attention with 99\% of the units installed after 2002 while 42\% of the fossil fuel boilers were installed previous to 1992 \citep{fleiter2016mapping}.  The large increase in the penetration of heat pumps in European homes has been motivated by various policy and regulatory drivers such as subsidies and carbon taxes, building regulations, improved technical standards and information dissemination \citep{hanna2016best, zimny2015polish}. \\

\noindent We focus on the European aggregated space heat demand and examine the extent to which it changes under the impact of global warming. We then estimate the consequent changes in the CO$_2$-emissions. To isolate the effect of climate change, we assume an unchanged stock of national heat generating technologies throughout the 21st Century. For each 144 km$^2$ grid cell, defined by the spatial resolution of the climate data, we map a cost-optimal technology for a historical time frame 1970--1990 and for an end-century time frame 2080--2100, and compare the differences. The demand and supply sides are modelled as highly temperature dependent. The heat demand is modelled through the heat load factor, which most commonly is determined as the fraction of the yearly averaged heat demand to the maximum. For the supply side, we introduce a simplified techno-economical standpoint of heat generation. Initially, we exclude policy decisions but a section is devoted to a policy assessment of the results. The application of state-of-the-art technical procedures combined with the large ensemble of highly granular climate data, that support the analyses, is considered as novel to existing literature. In summary, this approach provides new and more robust results that quantify the change in space heat demand throughout this century. \\

\noindent A limited amount of research has been devoted to this field, all with a focus on historical heating systems. \cite{frederiksen2013district} calculate the heat load capacity factors for 80 locations in Europe for a historical time frame spanning the years 1981--2000. \cite{scoccia2018absorption} compares, under historical weather conditions that are typical to the European region, the seasonal performances of six heating system configurations and finds results that are sensitive to the selection of electricity and gas driven heat pumps. \\ 

\noindent This paper is structured as the following: Section 2 presents the methodologies of this paper. Results are presented in Section 3. Current policies on decentralised heating in Europe along with future prospects are presented in Section 4 along with the study limitations. Conclusions and policy implications are presented in Section 5. Finally, a nomenclature is added in Section 6. 
\section{Methodology}

\noindent We devote this section to a qualitative description of the methods that are used in this work. A detailed derivation of the mathematical formulations can be found in SI 1.1 to 1.3.

\subsection{Technology and price assumptions}
\label{Sec: tech&price}
\noindent Inspired by Fig. \ref{Fig: tech_share}, the following listed technologies compose the ensemble of the decentralised heat generating technologies in this study: \\

\noindent \textbf{Electricity driven air source heat pumps (ASHP)} draw heat from the ambient air to supply hot water and space heating through hydraulic based water systems. Air source heat pumps require only an outdoor and indoor unit and are therefore easy to retrofit into existing houses. A limited amount of equipment and installation procedures give this technology an economic advantage compared to ground source heat pumps. On the other hand, the large temperature fluctuations between the external heat collector (source temperature) and the output at home (sink temperature) throughout the year, especially in winter periods with high heat demand and low ambient temperatures, challenge their efficiency, denoted by the Coefficient Of Performance (COP). In this work, we calculate temporally- and spatially-explicit COP values based on the prevailing air and soil temperature, with a sink temperature of 55 $\degree$C \cite{staffell2012review}. Full details are given in SI 1.3.\\

\noindent \textbf{Electricity driven ground source heat pumps (GSHP)} are identical to the previous, but draws heat from the soil instead, which offers a substantially higher yearly averaged COP. Temperature measurements from existing boreholes in Denmark show that at a depth of 20 meters, the ground temperatures have settled, i.e., become seasonally independent \citep{danish_bore}. In this work, we calculate the ground temperatures as an average of 20-year air temperatures. The resulting values correspond to temperatures at a depth of approximately 50 meters below ground, depending on soil type and geographical location \citep{danish_bore}. The higher capital investments of ground source heat pumps are compensated by the lower running costs compared to air source heat pumps. \\

\noindent \textbf{Air-to-air heat pumps with auxiliary electricity driven boilers (A2A+EB)} is a hybrid system that consists of an electric boiler for hot water supply and an air-to-air heat pump that draws heat from the ambient air and supply heat through air exchangers. Air-to-air heat pumps have the lowest capital investments of all the heat pumps. Furthermore, air-to-air heat pumps utilise a lower sink temperature, which in this work is assumed to be 30 $\degree$C \citep{staffell2012review}. This increases the COP further, and consequently reduces the running costs by around 20\% when compared to underfloor heating operating at 40 $\degree$C which is typical for GSHP \citep{staffell2012review}. Since air-to-air heat pumps cannot provide hot water they have to be installed alongside a hot water technology, which we assume is an electricity driven boiler. The combined technology efficiency will then be reduced. The share of each technology is determined by the individual shares of space heat and hot water demand to the total amount. \\

\noindent \textbf{Natural gas fired boilers} are assumed to cover both the hot water demand and space heat with hot water circulating through radiators. This technology has a very low capital cost but a relatively high running cost. \\

\noindent \textbf{Oil fired boilers} are identical to the previous but fired with oil. \\

\noindent \textbf{Biomass boilers} cover both the space heat demand and hot water by connection to radiators. The boiler is assumed to be automatically stocked. These types of boilers most commonly utilise wood pellets as fuel, which leads to the expensive fuel price in Tab. \ref{Tab: tech_summary}. \\

\noindent The stock of coal fired boilers has reduced heavily since 1992 with 58\% of all units being installed before 1992 and 12\% after 2002 \citep{fleiter2016mapping}. Currently, coal fired boilers comprise only 2\% of the total heating technology stock in Europe \citep{fleiter2016mapping}. The decreasing trend is mainly believed to be a result of aggressive CO$_2$ and air quality policies in the European countries. As a result, we exclude the coal fired boilers in this study. \\

\noindent Technologies such as fossil fuel driven boilers are very mature and possess relatively small price variability. Technologies such as heat pumps are still relatively new to the market and therefore subject to significant price variability between manufacturers and countries, and uncertainties in the future cost reductions and learning rates. These are mostly related to overcoming technological barriers, future markets and the technology demand \citep{ENdatasheet}. Upper and lower bounds of these uncertainties are presented in Tab. \ref{Tab: tech_summary} for all technologies that are included in this study. \\

\noindent In Tab. \ref{Tab: tech_summary} we summarise the technology properties and prices for retrofit into existing single family houses. As the focus of this paper is on the impact of climate change on heating across a continent, rather than modelling the bespoke heating mix in individual countries, all prices are excluding national taxes and levies, and assumed to be constant across regions. This allows for a direct measure and comparison of the impact of climate change across regions and time. Furthermore, this procedure reduces uncertainties related to national policies on tax regulations. For the same reasons, we ignore infrastructure constraints such as the absence of gas distribution networks in many countries and the inability of electricity distribution networks to meet large demands for heating from other countries, which is discussed further in Section \ref{Sec: stud_lim}. 

\begin{table*}[h!]
\caption{Technology costs and properties. The \emph{unperturbed} pricing scheme consists of installation, equipment and maintenance costs along with the uncertainty ranges that are prepared from Energinet \citep{ENdatasheet}. Since Energinet does not provide an underlying distribution for each uncertainty range, we take a conservative approach and assume that all installation, equipment and maintenance costs are uniformly distributed. This means that these expenses are equally likely to occur within a specific uncertainty range. Electricity and gas prices excluding taxes and levies are prepared from the Eurostat database \citep{eurostatel, eurostatgas}. Oil prices are prepared from the IEA database \citep{IEAoil}. Biomass prices (wood pellets) are prepared from the Cross Border Bioenergy project \citep{biomass}. The uncertainty range of fuel costs, $\sigma$, is defined according to a Gaussian distribution with a spread of 20\% of the fuel price as, e.g., in \cite{dahl2019cost}. A price variation of about $\pm$20\% is well represented by the coefficient of variation seen across annual-average gas and power prices in Europe over the last two decades \citep{eurostatel, eurostatgas}. Technology properties are prepared from Energinet \citep{ENdatasheet}. "t" denotes that the efficiency of heat pumps (COP) is temperature dependent.}

\scriptsize
\renewcommand{\arraystretch}{1.3}
\begin{tabular}{|p{3.5cm}|p{1.7cm}p{1.7cm}p{1.85cm}|p{1.7cm}p{1.7cm}p{1.8cm}|p{1.7cm}|} \hline
{} & \multicolumn{3}{c|}{\textbf{Boilers}} & \multicolumn{3}{c|}{\textbf{Heat pumps}} & \textbf{Biomass stoves} \\
{} & Gas Fired & Oil Fired & Electricity Driven & Air to Air (A2A) & Air to water (ASHP) & Ground to Water (GSHP) & {} \\\hline 
\textbf{Unperturbed pricing scheme} & {} & {} & {} & {} & {} & {} & {} \\
Installation Cost [\euro/kW] & 100 $\left[93,148\right]$ & 100 $\left[80,140\right]$& 50 $\left[30,70\right]$ & 75 $\left[50,83\right]$ & 304 $\left[240,480\right]$ & 420 $\left[350,560\right]$& 118 $\left[40,200\right]$ \\
Equipment Cost [\euro/kW] & 170 $\left[157,252\right]$ & 230 $\left[187,326\right]$& 50 $\left[30,70\right]$ &225 $\left[150,250\right]$ & 456 $\left[360,720\right]$ & 780 $\left[650,1040\right]$& 472 $\left[160,800\right]$ \\
Maintenance Cost [\euro/kW/yr] & 17 $\left[14,22\right]$ & 14 $\left[13,18\right]$& 7 $\left[5,10\right]$ & 22 $\left[17,25\right]$ & 24 $\left[19,30\right]$ & 24 $\left[19,30\right]$ & 25 $\left[16,27\right]$ \\ 
Fuel Cost [\euro/MWh] &        45  $\pm$ $\sigma$ & 64 $\pm$ $\sigma$ & 127 $\pm$ $\sigma$ & 127 $\pm$ $\sigma$ & 127 $\pm$ $\sigma$ & 127 $\pm$ $\sigma$ & 51 $\pm$ $\sigma$ \\ 
{} & {} & {} & {} & {} & {} & {} & {} \\
\textbf{Technology properties} & {} & {} & {} & {} & {} & {} & {} \\
Installed Capacity [kW] &        10 & 10 & 10 & 10 & 10 & 10 & 10 \\ 
Lifetime [yr] &                  20 & 20 & 20 & 12 & 18 & 20 & 20 \\ 
Discount rate [\%] &             4 & 4 & 4 &4 & 4 & 4 & 4 \\
Efficiency [\%] &                97 & 95 & 100 & t & t & t & 88 \\ 
\hline
\end{tabular}
\renewcommand{\arraystretch}{1}
\label{Tab: tech_summary}
\end{table*}

\subsection{Heat load factors}

\noindent The heat load factor, $\text{HLF}$, denoted as $\mu$, is defined as the unitless ratio of the residential heat demand, $L^{\text{Total}}$, to the maximum possible output of heat, $P^\text{Total}$, over a given time period, $\Delta$, as:

\begin{equation}
\mu = \frac{L^{\text{Total}}}{P^\text{Total} \cdot \Delta}
\label{Eq: mu}
\end{equation}

\noindent The decentralised nature of heating means that data on consumption is not readily available and therefore not applicable. Known to the literature, the theory of heating degree-days is most frequently used as a best proxy for modelling the variations in the day-to-day heat demand. The heating degree-days are calculated by using national temperature profiles. In this study, we average the 3-hourly temperatures into daily values, to emulate a night storage since the daily averaged value is higher than night-temperatures but lower than day-temperatures. The theory of the heating degree-days and its application to approximate $L^{\text{Total}}$ is described formally in SI 1.1. The maximum output of heat, $P^\text{Total}$, depends on the cold extreme temperature, as described in SI 1.1. \\

\noindent As stated previously, the heat load factors are determined as the fraction of the yearly averaged heat demand to the peak. Thus, large heat load factors are common in cold climates due to long running hours, but also in mild climates where hot water demand dominates the heat load. High load factors are consistent with a reduction in the overall cost per kWh of heat generated, since fixed expenses would be spread across more units of energy generated, hence the cost per unit of generation is reduced. Technologies with low marginal costs such as heat pumps prove as economically favourable in these circumstances. On the other hand, warmer climates tend to decrease the heat load factors, as peak hours deviate considerably relative to the base load hours. Technologies with low capital investments would serve as economically favourable in these conditions. 

\subsection{Techno-economic standpoint of heat generation}

\noindent In a simplified approach to the economics of heat generation, only the most significant costs and properties of decentralised technologies are included. Firstly, we include capital investments which are fixed, one time expenses, made up of equipment and installation costs. Equipment expenses cover the machinery including environmental facilities, whereas installation expenses cover engineering, civil works, buildings, grid connection, installation and commissioning of equipment \citep{ENdatasheet}. A yearly fixed maintenance expense is added. It includes all costs that are independent of how the technology is operated \citep{ENdatasheet}. Finally, we include marginal costs, which primarily depend on the technology operation time. Small scale effects such as the decrease in efficiencies or increase in maintenance expenses as a function of time are ignored. All technology properties and prices are formally introduced in Section \ref{Sec: tech&price}.\\

\noindent The hourly accumulated cost, $X^{\text{TOT}}_{\text{x}, \theta}$, for a technology, $\theta$, depends linearly on the heat load factor, $\mu_\text{x}$, at each grid location, x, as: 

\begin{equation}
\text{X}^{\text{TOT}}_{\text{x}, \theta} = \text{X}_{\theta}^{\text{CAP}} + \mu_\text{x} \cdot \text{X}_{\text{x},\theta}^{\text{OP}}
\label{Eq: Tot_cost}
\end{equation}

\noindent A detailed review of the capital and marginal expenses, $\mathrm{X}_{\theta}^{\text{CAP}}$ and $\text{X}_{\text{x}, \theta}^{\text{OP}}$, respectively, is conducted in SI 1.2. In this work, we have chosen to set the installed capacity to 10 kW for all technologies to scale the total cost to an appropriate value for a typical household. The capacity is kept fixed throughout the grid cells. However, the choice of capacity will not affect the results of this study as it is chosen to be identical for all technologies, see SI 1.2 for further details.

\subsection{Example of an application}
\label{Sec: Application}
\begin{figure}
    \centering
        \includegraphics[width=0.5\textwidth]{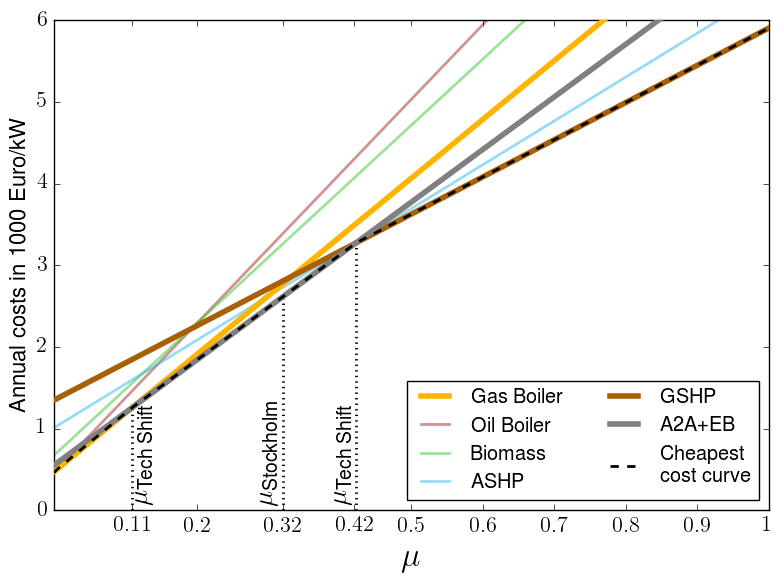}
        \caption{A screening curve showing annual cumulative heating costs in 1000 Euro/kW as a function of heat load factors, $\mu$, for the grid cell of southern Stockholm. $\mu_\text{Tech Shift}$ defines the heat load factor for which two technology crossing points occur. $\mu _\text{Stockholm}$ defines the heat load factor for Stockholm, Sweden.} 
    \label{Fig: cost_curves}
\end{figure}

\noindent The heat load factor, $\mu _x$, is calculated at first for each grid location, $x$, as shown in Eq. \ref{Eq: mu}. The heating expenses are then calculated for each technology, $\theta$, and each grid location, $x$, by using Eq. \ref{Eq: Tot_cost}. The data covers a modelling time span of 20 years, since 20 years define the approximate extent of a climatic period and the typical lifespan of a heating technology.  Fig. \ref{Fig: cost_curves} shows the accumulated expenses as a function of the heat load factor, $\mu$, for the grid cell of southern Stockholm, Sweden. This is termed in the screening curve for heating technologies, analogous to the screening curves used for comparing electricity generation costs \citep{staffell2015there}. In the case of oil and biomass boilers, high fuel prices and low efficiencies result in large operational expenses, which make these technologies highly uncompetitive. On the other hand, high COPs of heat pumps compensate for their high capital investments, which make these technologies competitive to gas boilers at high values of $\mu$. \\

\noindent For Stockholm, the optimal technologies consist of gas boilers, air-to-air heat pumps with auxiliary electricity driven boilers and ground source heat pumps. For heat load factors below $0.11$, gas boilers would serve as the cost-optimal choice for heating purposes. Since heat load factors never reach this domain it stays as non-applicable. For heat load factors between $0.11$ and $0.42$, air-to-air heat pumps with auxiliary electricity driven boilers would serve as a cost-optimal choice. Finally, for heat load factors above $0.42$, ground source heat pumps would be cost-optimal. The heat load factor of southern Stockholm, $\mu_\text{Stockholm}$, equals 0.32, for which air-to-air heat pumps with auxiliary electricity driven boilers would serve as cost-optimal. This procedure is repeated for each of the 412 x 424 grid locations in the data set and for all of the nine climate models. \\

\section{Results}
\label{Sec: Results}
\subsection{Impact of climate change on the heating degree-days}
\begin{figure}
    \centering
        \includegraphics[width=0.5\textwidth]{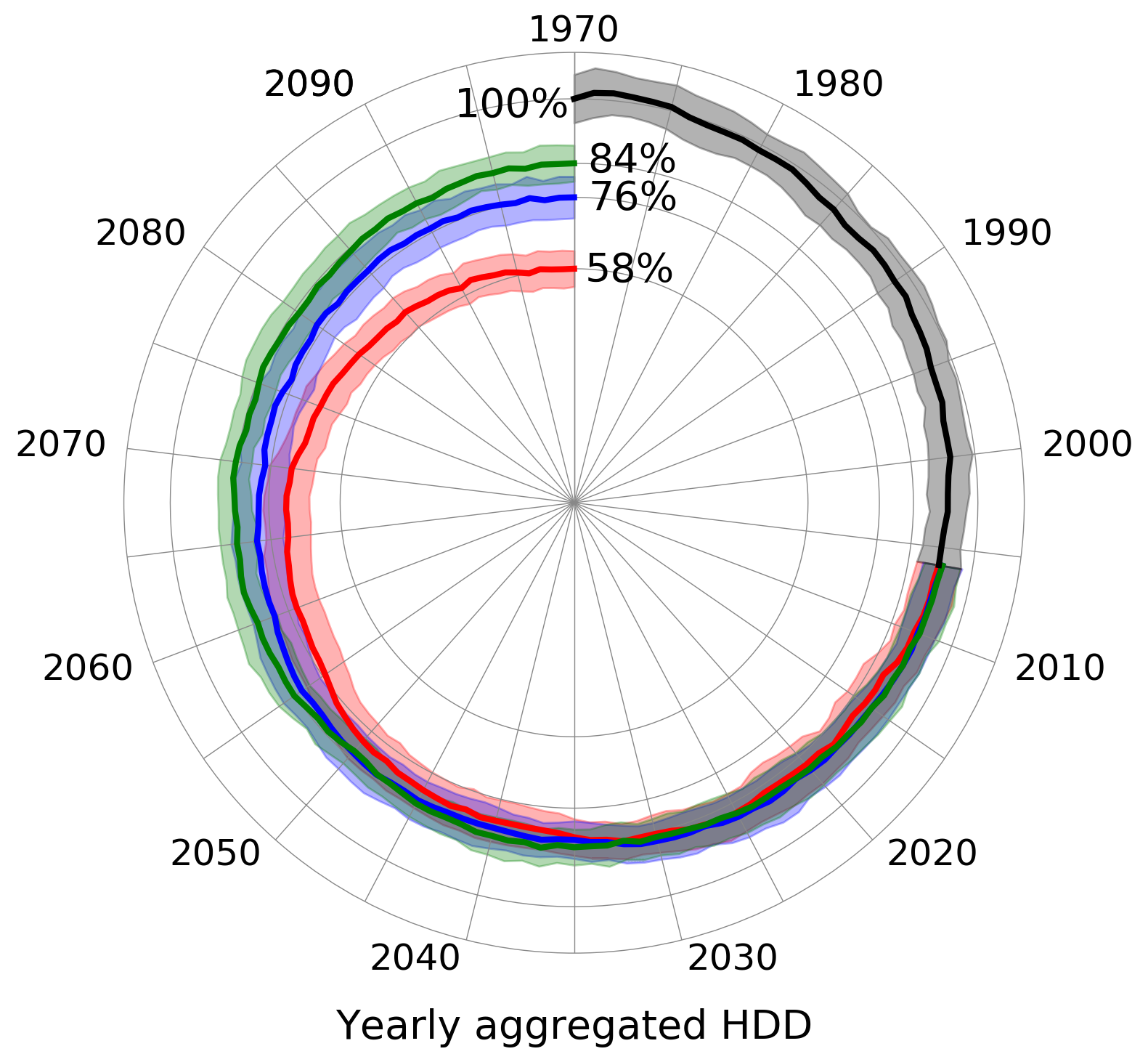}
        \caption{10 year moving average of the yearly aggregated heating degree-days (full drawn curves) for each of the three projections of climatic outcomes, RCP2.6 (green), RCP4.5 (blue) and RCP8.5 (red). Each yearly result is composed of a climate model ensemble average and shown in percent of the corresponding 1970 value. $\pm 1\sigma$ standard deviations are shown with shaded regions.} 
    \label{Fig: HDD_vs_year}
\end{figure}

\noindent Initially, we present results that show the extent to which the European heating degree-days change under the impact of global warming. The gridded temperature profiles, $T_\text{x}(t)$, have been weighted according to the population density in each grid cell \citep{gridded_pop}. The population weighted temperature profiles are used to calculate the heating degree-days, which have been aggregated by summing over all grid-cell values within a country. The weighting is especially important for the Nordic countries as, e.g, Norway, where the sparsely populated areas in the north, otherwise, would contribute significantly to the aggregation. Fig. \ref{Fig: HDD_vs_year} presents the yearly aggregated heating degree-days for Europe from 1970 to 2100 for each of the three projections of climatic outcomes, RCP2.6, RCP4.5 and RCP8.5. Each yearly result is composed of a climate model ensemble average and shown relative to the corresponding 1970 value. A 10 year moving average (full drawn curves) is used to highlight the long-term trends over annual fluctuations. It is clear that all climate change pathways result in a decreasing trend in the heating degree-days, with magnitudes being specific to the climate conditions of each RCP. The year of 2100 in RCP8.5 shows a decrease of approximately 42\% in comparison to 1970, which is a consequence of almost 5 $\degree$C temperature increase in the business-as-usual scenario. Corresponding values for RCP2.6 and RCP4.5, are 16\% and 24\%, respectively. The uncertainties stay below $\pm$8\% of the ensemble average for all RCPs, which provides a strong evidence of agreement among the ensemble of climate models. The temperature data have been bias adjusted as explained in SI 1.4. \\

\noindent To estimate the resulting change in the CO$_2$-emissions from space heating, we assume a fixed national stock of heating technologies throughout the century, to remove the effect of technological change and to isolate the effect of climate change. Based on 2015 values, the production of electricity and heat in the EU28 accounted for approximately 30\% of total CO$_2$-emissions, with heat production accounting for more than half of this share \citep{iea2}. Altogether, a decrease of 42\% in the heating degree-days for RCP8.5, leads to a decrease of 12.5\% in the CO$_2$-emissions. For RCP2.6 and RCP4.5 the respective values are 4.8\% and 7.2\%. In the following, we focus on the supply side of heat, in a search for the cost-optimal technologies to cover the changing demand throughout the 21st Century. \\

\noindent For the remainder of the results section, we move way from the aggregated approach and instead focus on each grid location separately. For each location, we determine the cost-optimal heating technology by following the procedure explained in Section \ref{Sec: Application}. Initially, we present results only based on a predefined reference period for the ICHEC-EC-EARTH HIRHAM5 climate model. We then present equivalent results for all climate projections and climate models, but we stress, however, that during the analysis, all data has been treated equivalently. \\

\subsection{Cost-optimal heating technologies in a historic time frame}

\noindent The unperturbed cost assumptions, presented in the first four rows of Tab. \ref{Tab: tech_summary}, are subjected to uncertainties that strongly reflect the maturity of the technologies. These will naturally propagate into output uncertainties, meaning that the selection of cost-optimal technologies might be as uncertain as the input prices, which they are subjected to. In order to assess the robustness of the selection of cost-optimal technologies, the optimisation process, as explained in Section \ref{Sec: Application}, has been run with 100 Monte Carlo trials for the pricing scheme. Each pricing scheme consists of a random perturbation of the unperturbed installation, equipment, maintenance and fuel prices subjected to their respective uncertainty ranges. \\

\begin{figure}
\centering
\begin{subfigure}[b]{0.45\textwidth}
   \includegraphics[width=1\linewidth]{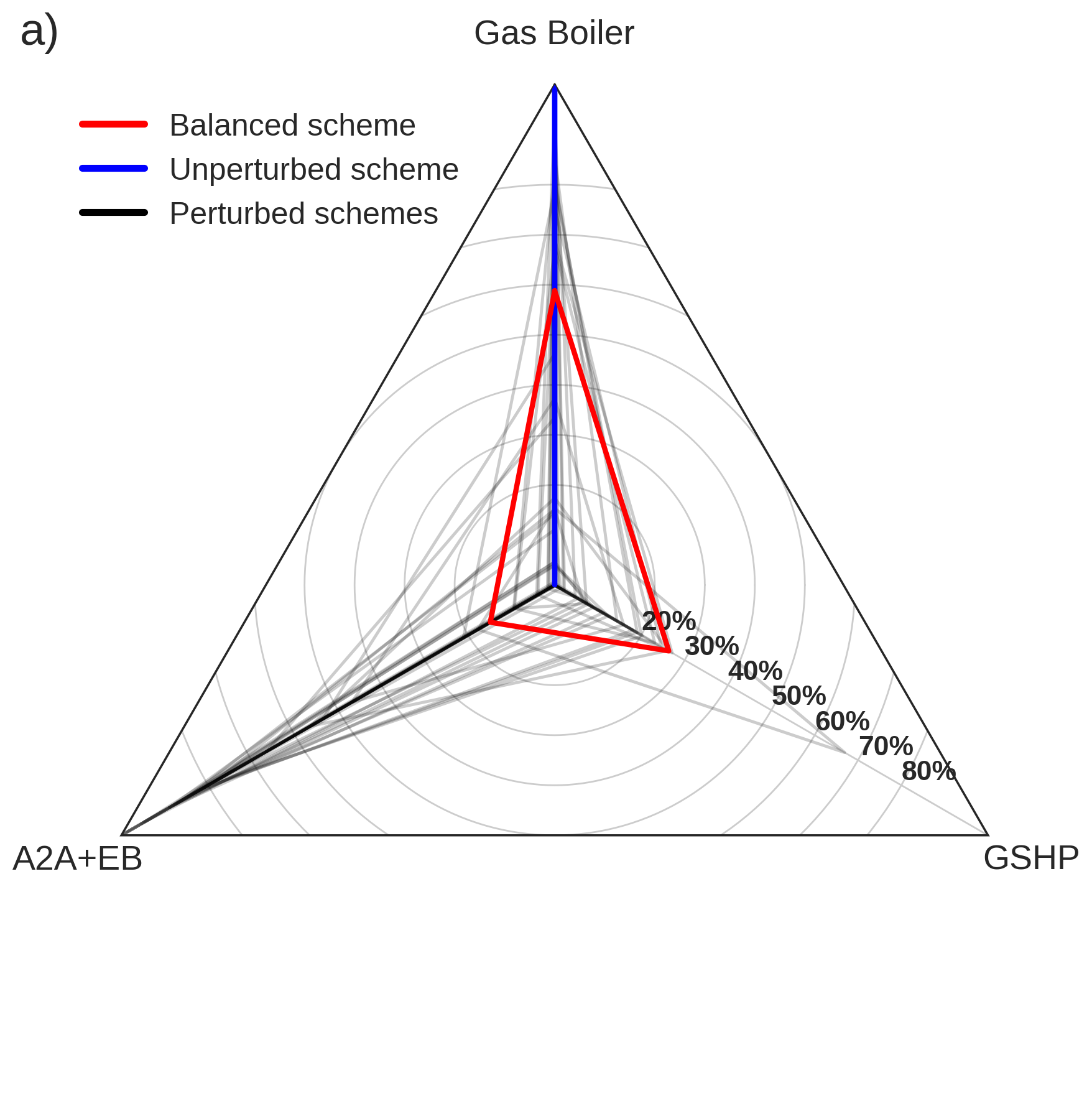}
\end{subfigure}
\begin{subfigure}[b]{0.45\textwidth}
   \includegraphics[width=1\linewidth]{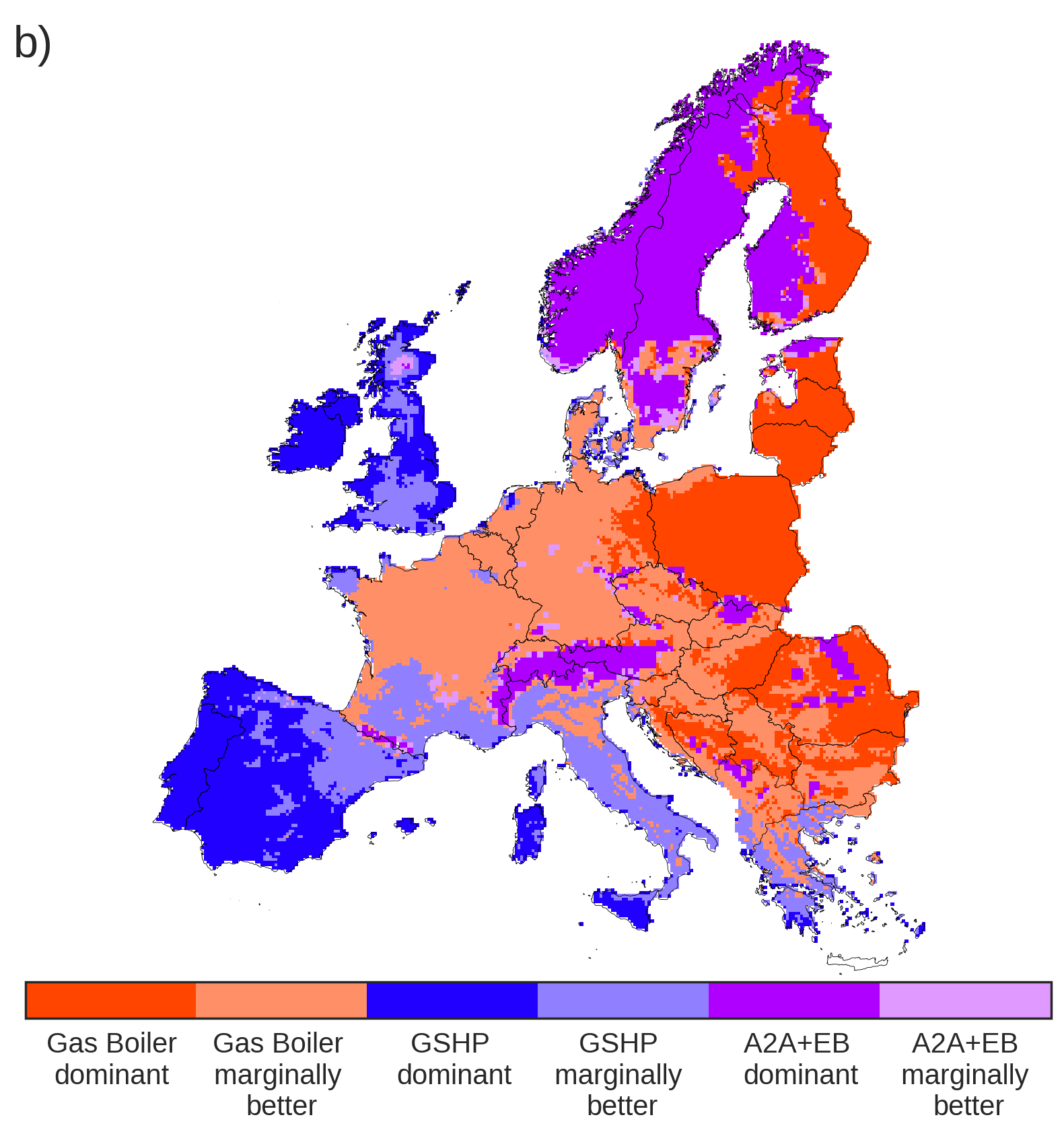}
\end{subfigure}
\caption{Panel a: Cost-optimal technology distributions based on the ICHEC-EC-EARTH HIRHAM5 climate model. Blue line is a result of the \emph{unperturbed} pricing scheme. Shown with red is the resulting technology distribution by using the \emph{balanced} pricing scheme. Black curves show technology distributions from 99 perturbed pricing schemes (see Eq. \ref{Eq: pricing_sel}). Panel b: Spatial distribution of technologies resulting from the \emph{balanced} pricing scheme, which is summarised with red in Panel a. GSHP denotes ground source heat pumps. A2A+EB denotes the hybrid of air-to-air heat pumps with auxiliary electricity driven boilers.}
\label{Fig: unc_spa}
\end{figure}

\noindent The results of the optimisation processes are presented on the radar chart in Fig. \ref{Fig: unc_spa}a for all pricing schemes. Each plot shows the normalised number of grid cells (proportional to land area across Europe) for which a technology serves as a cost-optimal, i.e., sum of the technology shares for a single plot is 100\%. It becomes clear that each individual pricing scheme defines a unique technology distribution, resulting in a highly cost-sensitive outcome of the optimisation. The \emph{unperturbed} pricing scheme, as presented in Tab. \ref{Tab: tech_summary}, has a tendency towards single technology dominance, i.e., gas boilers serve as cost-optimal in all grid locations, as shown with blue in Fig. \ref{Fig: unc_spa}a. This results from a combined effect of the range of possible fuel prices, which is larger than the diversity in heat load across grid cells and the substantially lower gas boiler prices compared to the costs of the remaining technologies. We provide a detailed discussion on the single technology dominance in SI 2.1. Focusing briefly on the end of century climatic periods, we find that the \emph{unperturbed} pricing scheme leads to an identical technology distribution as for the historical period, for all projections of climatic outcomes. This is an important issue to address, as with this pricing scheme, the impact of climate change will not show its potential significance in the selection of technologies. The significance of climate change is therefore clarified by selecting a pricing scheme that to a high degree defines a balanced distributions of technologies. The selection is based on minimising the sum of squares of the difference between a technology share, $\theta_i$, and the maximum appearing technology share, $\theta_{\text{max},i}$, as shown in Eq. \ref{Eq: pricing_sel}, which leads to the \emph{balanced} pricing scheme, presented in Tab. \ref{Tab: balanced_summary}. The \emph{balanced} pricing scheme is classified as a unique outcome of the perturbation process, where the gas boiler expenses increase significantly compared to the respective increase in heat pump expenses, which gives heat pumps an economic benefit. The red plot in Fig. \ref{Fig: unc_spa}a illustrates the resulting technology share by using the \emph{balanced} pricing scheme. Gas boilers cover 59\% of Europe whereas ground source heat pumps cover 26\% and the hybrid of air-to-air heat pumps and auxiliary electricity driven boilers cover 15\%. 

\begin{equation}
\min_{\theta} \sum_i \left(\theta_i - \theta_{\text{max},i}\right)^2
\label{Eq: pricing_sel}
\end{equation}

\begin{table*}[h!]
\caption{The \emph{balanced} pricing scheme, which is designed to enforce a balanced distribution of technologies across Europe. The technology properties remain unchanged during the perturbation process, as presented in Tab. \ref{Tab: tech_summary}.}

\scriptsize
\renewcommand{\arraystretch}{1.3}
\begin{tabular}{|p{3.5cm}|p{1.1cm}p{1.1cm}p{1.2cm}|p{1.7cm}p{1.7cm}p{1.8cm}|p{1.5cm}|} \hline
{} & \multicolumn{3}{c|}{\textbf{Boilers}} & \multicolumn{3}{c|}{\textbf{Heat pumps}} & \textbf{Biomass stoves} \\
{} & Gas Fired & Oil Fired & Electricity Driven & Air to Air (A2A) & Air to water (ASHP) & Ground to Water (GSHP) & {} \\\hline 
\textbf{Balanced pricing scheme} & {} & {} & {} & {} & {} & {} & {} \\
Installation Cost [\euro/kW] & 117  & 99 & 64 & 65 & 347 & 443 & 84  \\
Equipment Cost [\euro/kW] & 200 & 293 & 48 & 194 & 520 & 824 & 336 \\
Maintenance Cost [\euro/kW/yr] & 18 & 17 & 7 & 21 & 24 & 24 & 23 \\ 
Fuel Cost [\euro/MWh] &        65 & 80 & 144 & 144 & 144 & 144 & 59 \\ 
\hline
\end{tabular}
\renewcommand{\arraystretch}{1}
\label{Tab: balanced_summary}
\end{table*}

\noindent Oil and biomass boilers, and air source heat pumps are not economically viable for all pricing schemes and therefore not considered further. For most of these technologies, this is explained by a combination of high oil, biomass and electricity prices and low technology efficiencies. The unexpected outcome that air source heat pumps serve as economically unfavourable, can be justified by the technology COP that to a high degree determines the operational expenses. The empirical equations for the COPs, as presented in (SI.3), are highly dependent on the heat source and sink temperatures. Soil temperatures are in general higher and more uniform over the the winter heating season compared to air temperatures. Ground source heat pumps therefore offer higher COPs over the year compared to air source heat pumps for any given sink temperature. Again, this reflects purely techno-economic potential, and does not consider geological or social barriers to uptake, especially the significant disruption caused by retrofit installations \citep{staffell2012review}. Depending on the hot water to the space heat ratio, the combined technology efficiency of air-to-air heat pumps with auxiliary electricity driven boilers will lower accordingly. \\

\noindent The spatial distribution of the cost-optimal technologies, resulting from the \emph{balanced} pricing scheme, is shown in Fig. \ref{Fig: unc_spa}b. This reflects where technologies would be best placed throughout Europe if there were homogenous prices, policies and public attitudes towards each - which is not the case in reality.  The difference between the present-day distribution of technologies and this figure shows the impact of non techno-economic considerations on heating choice. The "marginally better" category defines a $\pm$5\% region around the intersection point of two cost curves, which defines an indecisive region where either of the technologies can provide a cost-optimal option for heating. The "dominant" category defines the outside region. Each technology is therefore subjected to one of the two categories: 

\begin{align*}
&\text{Marginally better:} \\
&\mu_\text{Tech Shift, x} -0.05 < \mu_x < \mu_\text{Tech Shift, x} + 0.05 \\\\
&\text{Dominant:}\\
&\mu_\text{Tech Shift, x} - 0.05 \geq \mu_x \\
&\mu_\text{Tech Shift, x} + 0.05 \leq \mu_x\\
\end{align*}

\noindent where $\mu_\text{Tech Shift, x}$ defines the heat load factor at the intersection point of two cost curves at a grid location, x, as shown in Fig. \ref{Fig: cost_curves}. Interestingly, the historical time frame does not illustrate an expected north-south dipole, but a split for which heat pumps are dominating the coastal areas, and gas boilers the mainland areas. This result is partially explained by the Köppen--Geiger climate classification system for Europe \citep{geiger1954klassifikation} and partially by the assumption of constant hot water use across Europe, as detailed in the following. \\

\noindent The cold oceanic climate of West Europe results naturally in a large heat consumption throughout the year that, as a result, increases the heat load factor to a value of 0.5, depending on the grid location. Ground source heat pumps serve as cost-optimal in these regions, as a result of their relatively low operational expenses at high heat load factors. Air-to-air heat pumps with auxiliary electricity driven boilers serve as cost-optimal in Scandinavia. This is a result of the low hot water to space heat ratio in cold climates, which in turn limits the decrease of the combined technology efficiency. Contrary to North-West Europe, the tempered Mediterranean climate results in a decreased energy consumption for space heating. The hot water to the space heat ratio increases therefore significantly, which results in increased heat load factors. As for the West European areas, technologies with low operational expenses for high heat load factors serve as economically favourable. On the other hand the increased hot water to space heat ratio, decreases the combined efficiency of the air-to-air heat pumps with auxiliary electricity driven boilers, which makes this technology economically unfavourable in temperate climates. The overall cold winters and hot summers in the East European mainland result in low heat load factors. As a result, gas boilers become economically favourable. \\

\noindent From this point, the paper will only discuss results that are based on this pricing scheme, which we refer to as the \textit{balanced} pricing scheme. \\

\subsection{End-of-century projections}
\noindent Fig. \ref{Fig: best_tech_basemap} shows the spatial distribution of cost-optimal technologies resulting from the \textit{balanced} pricing scheme. These are now shown for the three end-of-century time periods defined to span the years 2080--2100 for all of the climate projections. Historical results are added for easy reference. The most striking observation to emerge is the large increase in the attractiveness of heat pumps towards continental Europe, as a result of changing climatic conditions. This is the product of many interlinked factors, considering that climate change affects both the supply and demand side simultaneously. However, the main effect is observed on the supply side, because of increasing heat pump COPs due to increasing winter temperatures. Another contributing factor might be the increasing heat load factors that emerge as a result of a fixed hot water use and a decreasing demand for space heat. For Scandinavia, the hot water to space heat ratio increases significantly due to a decreased need for space heating. As a result, the combined efficiency of air-to-air heat pumps with auxiliary electricity driven boilers decreases and therefore makes this technology economically unfavourable and out-priced by gas boilers. \\

\subsection{Climate model ensemble average}
\begin{figure*}
    \centering
        \includegraphics[width=0.9\textwidth]{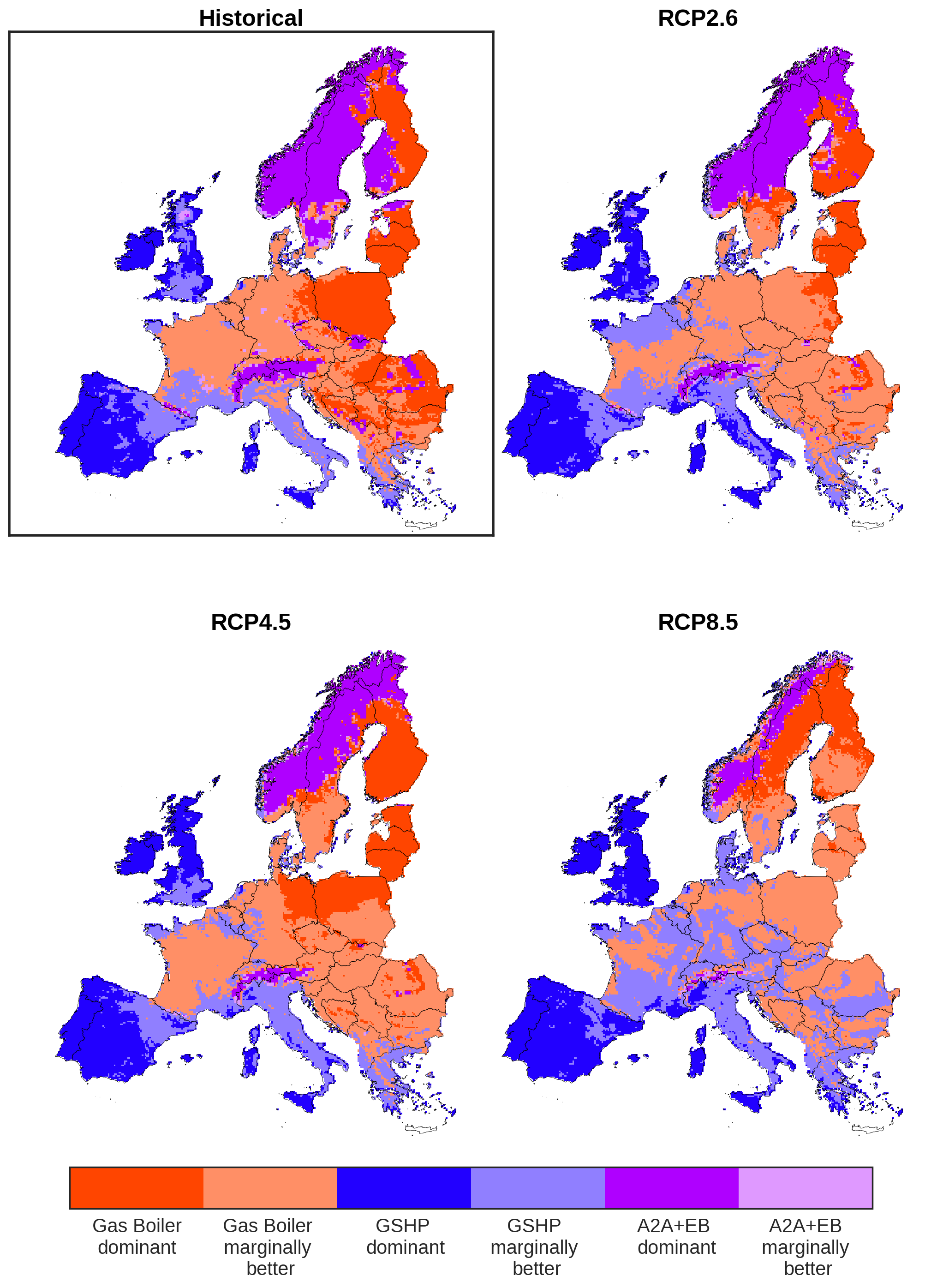}
        \caption{Spatial distributions of the cost-optimal heating technologies by using the \emph{balancing} pricing scheme. The historical period is defined to span the years 1970--1990. RCP2.6, RCP4.5 and RCP8.5 spans a climatic period from 2080--2100. GSHP denotes ground source heat pumps and A2A+EB denotes the hybrid technology of air-to-air heat pumps with auxiliary electricity driven boilers.} 
    \label{Fig: best_tech_basemap}
\end{figure*}

\begin{figure*}
    \centering
        \includegraphics[width=1\textwidth]{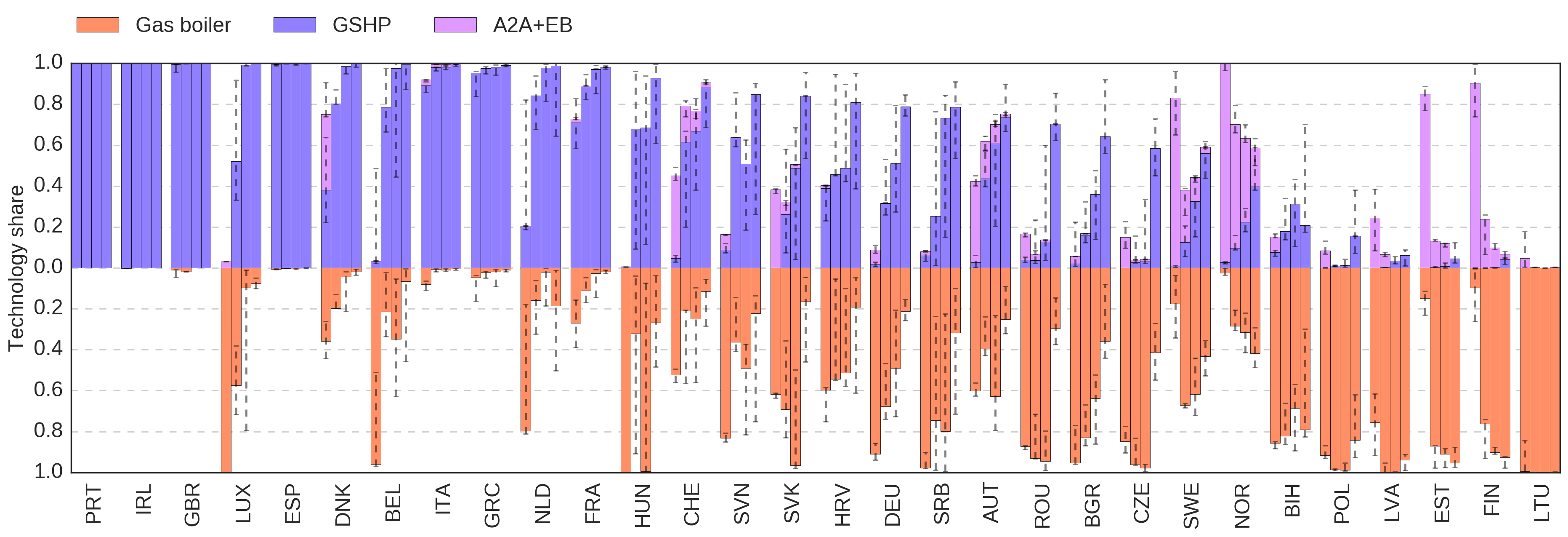}
        \caption{Climate model ensemble average of the population weighted shares of the national technology distribution. For each country the order of bars represent results from the historical, RCP2.6, RCP4.5 and RCP8.5 climate projections. Error bars illustrate the respective 25th and 75th percent quantiles. The shares of gas boilers and heat pumps have been separated for the matter of visualisation. The y-axis illustrates therefore the same quantity but in separate directions. GSHP denotes ground source heat pumps and A2A+EB denotes the hybrid of air-to-air heat pumps with auxiliary electricity driven boilers.} 
    \label{Fig: best_tech_bar}
\end{figure*}

\noindent Finally, in Fig. \ref{Fig: best_tech_bar} we summarise the results of Fig. \ref{Fig: best_tech_basemap} for all climate models. The bars illustrate the climate model ensemble average of the population weighted technology distributions, for each individual country. The categories of dominance, "marginally better" and "dominant" have been merged in this figure. The errorbars denote the respective 25th and 75th percentiles. The bars are ordered according to descending shares of ground source heat pumps in the RCP8.5 end-of-century time frame. It is clear that the heating infrastructure for the far west and far south countries is largely unaffected by the climate induced weather changes. This can be seen from the largely unaffected shares of ground source heat pumps for the different climate periods. The robustness of this result is confirmed by the small uncertainties, illustrating a high agreement among climate models. Norway and Sweden make a sharp transition from air-to-air heat pumps with auxiliary electricity driven boilers to a mixture of gas boilers and ground source heat pumps, also confirmed by the high agreement among the climate models. The Baltics, including Finland make a transition from air-to-air heat pumps with auxiliary electricity driven boilers towards strong gas boiler dominance with high agreement across models. For the remaining countries, it is clear that a higher degree of climate change, suggests a transition from gas boilers to ground source heat pumps. These results come with relatively large uncertainties, which results in a fluctuating degree of technology transition among the climate models. In general, it is observed that a higher degree of global warming tends to increase the stock of heat pumps towards the mainland of Europe. On the other hand, this trend is difficult to compare between the climate projections because of different underlying assumptions from the Integrated Assessment Modelling. 

\section{Discussion}
\label{Sec: pol_impl}
\subsection{Current status of heat policies in Europe affecting the stock of heating technologies}

\noindent This paper finds that across Europe as a whole, space heating demand declines by 16\% to 42\% under different climate change projections from 1970 to 2100. This is consistent with other studies, such as \cite{van2019amplification}, who observe decreased heating demand in Europe across an ensemble of Earth System Models under RCP4.5 and RCP8.5. Similarly, \cite{kitous2017assessment} find a 37\% decrease in residential heating demand by the period 2071-2100, compared to 2010, based on five global climate scenarios. In section \ref{Sec: Policy}, we consider general policy implications in particular relating to the flexibility of existing heat policy frameworks to adapt to reduced heating loads given more stringent building regulations and climate change. \\

\noindent The second part of our analysis points to different zones of Europe where heat pumps or gas boilers may be more or less optimal in cost or performance terms given projected temperature increases to the end of the century. In general, ground source heat pumps are shown to be more economically optimal in western and southern Europe, whereas gas boilers are more optimal in eastern Europe and some Nordic countries. \\

\noindent In sections \ref{Sec: Tech_south_east} to \ref{Sec: prospect}, we compare our numerical findings on cost-optimal decentralised heating technologies under climate change projections, with the current state of policies, national policy strategies and technological deployment in different European countries. We have modelled cost-optimal technologies for the period 2080-2100, and we argue that current heat policies and longer-term strategies for heat decarbonisation, for example to 2050, are relevant to the interpretation of these findings. Although the European stock of heating technologies will have undergone several replacement cycles by the late 21st Century, there are important sources of path dependency and lock-in that have led in particular to increasing returns to adoption of incumbent heating technologies such as natural gas boilers in the UK, or biomass-based district heating in Sweden \citep{gross2019path}. Without policy intervention to address these, there is a risk that existing, incumbent heating technologies and linked infrastructures are self-perpetuating, limiting opportunities for and slowing the deployment of alternative decentralised heating technologies such as heat pumps. In addition, it would be advantageous to support learning and cost reduction of heat pumps \citep{kiss2014heat} through policies to support their increased installation in regions where they are likely to become more cost-optimal under climate change in the longer term. \\

\noindent We find that in general, national heat policy outcomes and intentions align reasonably well with cost-optimal technologies as indicated by our model. However, there are also important mismatches between cost-optimal technologies and their real-world deployment. These mismatches are at least partly a product of the presence or absence and balance of policies which support or hinder the deployment of heat pumps or gas boilers. We argue that policy makers should be mindful of which technologies are most economically optimal in particular regions both currently and under projected climate change, in order to deliver cost-optimal policy outcomes. \\

\noindent The focus of sections \ref{Sec: Tech_south_east} to \ref{Sec: prospect} is on national-scale policies in order to understand variation between countries. Nevertheless, a number of EU heat pump support policies have been implemented and are interconnected with some national policies. For example, the EU Renewable Energy Directive in 2009 officially recognises heat pumps as a renewable energy technology \citep{EUParCoun}. It sets out a minimum efficiency level needed to produce renewable heat from electrically-driven heat pumps, equivalent to a seasonal performance factor (SPF) of over 2.88 \citep{gleeson2013meta}. This Directive also contributed to quality assurance in Europe by requesting that member states should introduce or have in place certification schemes for installers of heat pumps by 2012 \citep{EUParCoun, rizzi2011towards}. In 2016, the European Commission published the first EU Heating and Cooling Strategy \citep{EC2016}. This strategy was endorsed by the European Parliament, which also proposed that subsidies supporting fossil fuel heating should be phased out \citep{european2017european}. 

\subsection{Policy frameworks and changes in heating demand}
\label{Sec: Policy}

\noindent The findings of our study point to lower heating demand with increased global warming. We might also expect heating demand to fall further through the 21st Century based on EU-wide and national efforts to tighten building standards. Policies and regulations promoting building energy efficiency are relatively recent and will take time to address around three quarters of the European building stock which is considered to be energy inefficient \citep{honore2018decarbonisation}. There are significant challenges with reducing heat loss from buildings and replacing old, inefficient heating appliances, and these challenges vary across different sectors, e.g. service buildings, public buildings, problems with split incentives in privately-rented homes, industry \citep{EC2016, vivid}.  \\

\noindent In Europe, mandatory standards at an EU and national level have resulted in increased thermal efficiency of new and refurbished buildings, and greater use of more efficient heating appliances such as condensing boilers and heat pumps, leading to decreasing heat demand at an individual building level \citep{EEA2019, gynther2015energy, vivid}. There is significant potential for this trend to continue with further replacement and renovation of existing buildings, particularly since the average number of new dwellings built every year from 2000 to 2012 represented just 1\% of the EU housing stock \citep{gynther2015energy}. In 2016 in the EU, heating consumption per m$^2$ was 68\% of the level of heating demand in 1990. However, overall heating consumption only declined by 4\% due to the increase in building floor area over this period \citep{EEA2019}. \\
 
\noindent The Energy Performance of Buildings Directive was introduced in the EU in 2010 and stipulates minimum energy efficiency standards for building renovations, while also requiring new buildings to consume close to zero energy from the end of 2020 \citep{honore2018decarbonisation}. The Energy Efficiency Directive (2012) further requires EU member states to produce a long-term strategy for investing in improving the energy performance of existing residential and commercial buildings after 2020. This should include action plans for reducing heating and cooling demand and undertaking deep building renovations typified by 60\% or greater energy efficiency improvements. These regulations are technology neutral and therefore allow flexibility to achieve stipulated energy savings with a diversity of highly-efficient, decentralised heating technologies \citep{honore2018decarbonisation}. \\

\noindent Similar mandates have also been enacted at a national level in Europe. For example, Germany introduced the Energy Efficiency Ordinance to implement the European Energy Performance of Buildings Directive. In the Netherlands, progressively stronger energy efficiency standards have resulted in decreasing heat demand. France introduced the Régulation thermique in 2012, requiring all new buildings (constructed from 2013 onwards) to meet a maximum level of primary energy consumption meaning that direct electric heating could not be used in new buildings \citep{vivid}.

\subsection{Technology support policies: southern and eastern Europe}
\label{Sec: Tech_south_east}

\noindent The findings in Section \ref{Sec: Results} indicate that ground source heat pumps are favourable under the end-of-century scenario in southern European countries with a Mediterranean climate such as Spain, Portugal, Italy and Greece. However, there appears to be relatively limited policy support for heat pumps in these countries, where the majority of sales are for air-to-air heat pumps which can provide both heating and cooling \citep{european2014european, european2017european}. In Portugal, there has been an absence of national incentives for heat pumps \citep{european2014european}. In Spain, reversible air-to-air heat pumps are mainly used for cooling; gas boilers are more likely to be used for space heating \citep{european2015european}. Small air-to-air heat pumps may be combined with electric heating to provide both heating and cooling in some Mediterranean coastal zones of Spain. Italy has made tax reductions available for the installation of heat pumps on the condition that high seasonal performance factors are achieved \citep{economics2013pathways, hanna2016best}. There have also been information dissemination activities - the Gruppo Italiano Pompe di Carole is a group of manufacturers which promotes heat pumps through exhibitions, seminars and training courses \citep{european2015european}. \\

\noindent Most homes in Greece use diesel heating oil for space heating, while electricity, natural gas and biomass provide less than 12\% of total space heating each \citep{papakostas2015}. The installed capacity of ground source heat pumps in Greece has experienced a rapid growth in the last two decades, from 0.4 MW$_\text{thermal}$ in 1999 to 135 MW$_\text{thermal}$ in 2014 \citep{karytsas2017review}. This is a result of several factors, including rising oil prices compared to electricity prices, higher public awareness of ground source heat pumps and legislation introducing a licensing process for installations. In recent years, the development of the ground source heat pump market has been adversely affected by the economic crisis and an inactive construction industry \citep{karytsas2017review}. \\

\noindent The results show gas boilers to be optimal in eastern Europe particularly under RCP2.6 and RCP4.5, see Fig. \ref{Fig: best_tech_bar}. There is uneven agreement between this finding and the direction of heating policies in different eastern European countries. Following the decline of communism, many eastern and central European countries began to shift away from expensive and poorly maintained district heating systems towards individual household heating technologies \citep{poputoaia2010regulating, bouzarovski2016locked}. In Poland, coal is the dominant fuel for heating, and therefore a transition to gas heating and/or heat pumps would help to improve air quality and achieve decarbonisation. 40\% of Poland's 13 million houses use individual coal boilers or furnaces for space heating \citep{european2015european, european2017european}, while coal-fuelled district heating supplies space heating to approximately a further 30\% of total dwellings \citep{european2017european, wojdyga2017chances}. Around two thirds of the 2 million gas boilers installed in Polish houses are used only as a supplementary heat source in cold periods \citep{european2015european}. Despite a lack of specific policy support for heat pumps from the Polish state, sales of heat pumps grew to over 20.000 per annum by 2015 \citep{european2015european, european2017european, zimny2015polish}. \\
 
\noindent Heat pump sales in Slovakia have been constrained by the presence of a dense gas network \citep{european2015european}. There were no investment subsidies for heat pumps prior to 2015, and retrofitting heat pumps to buildings with existing gas connections is an unattractive investment; most market potential for heat pumps is in the new build sector. In the Czech Republic, our findings reveal ground source heat pumps as economically favourable in RCP8.5 (although gas boilers are more cost-optimal in the other projections). Investment subsidies were introduced in the Czech Republic in 2014 and support ground source and air source heat pumps which achieve a minimum efficiency standard \citep{european2015european}.
 
\subsection{Technology support policies: northern and western Europe}
\label{Sec: Tech_north_west}

\noindent The findings in this study show that ground source heat pumps are more cost-optimal in western Europe even under moderate temperature increases, but less optimal in northern Europe. In a number of western and northern European countries, a combination of policies have been effectively implemented to stimulate the take-up of heat pumps, including policies to raise technical standards, promotion and information campaigns and investment subsidies \citep{hanna2016best}. Up-front grants have been provided in Austria for consumers to switch from fossil fuel heating to heat pumps which achieve minimum performance standards based on the seasonal performance factor \citep{kranzl2013renewable}. In Denmark, an information campaign was followed by a subsidy scheme in 2010 to promote the replacement of dilapidated oil heaters with energy efficient heat pumps \citep{nyborg2015heat}. A subsidy scheme in Sweden from 2006 to 2010 made up-front grants available for households to switch from oil heating to heat pumps, district heating or biomass. The entire budget for this subsidy was used up after the first year of the scheme, with heat pumps being the most popular replacement for oil heating \citep{ericsson2009introduction, vivid}. \\

\noindent Long-term stability of policy support has been an important success factor for substantial deployment of heat pumps in countries such as Sweden, Switzerland and Austria, since this increases industry and consumer confidence \citep{hanna2016best}. Carbon and fuel taxes represent further instruments capable of incentivising the switch from fossil fuel heating to low carbon alternatives. Carbon taxes have been adopted, in particular, by northern European countries since the early 1990s, including Finland, Norway, Sweden and Denmark \citep{sumner2011carbon}. Sweden has the highest carbon tax in Europe, which has been increased threefold since 1991 \citep{WorldBank}. Separate taxes have also been applied on natural gas heating (although it has limited presence in Sweden) and heating oil. Energy and carbon taxes in Sweden have helped to encourage the substitution of oil boilers with heat pumps \citep{ericsson2009introduction, vivid}. \\

\noindent In Germany, where 54\% of households are connected to the gas grid \citep{hanna2016best}, there were over 700.000 heat pumps in operation in 2015 \citep{european2017european}. This is consistent with cost-optimal technologies indicated for this country in Fig. \ref{Fig: best_tech_bar}: principally gas boilers in RCP2.6 and RCP4.5, and a mixture of heat pumps and gas boilers in RCP8.5. Heat pumps have replaced some gas boilers in Germany, but this is largely restricted to new build homes: the share of heat pumps' contribution to primary heating energy increased from 1\% of new build homes in 2000 to 27\% in 2017. The equivalent share for natural gas declined from over three quarters of new homes in 2000 to 39\% in 2017 \citep{AGEB18}. \\
 
\noindent A combination of building regulations and investment subsidies have helped to increase the proportion of heat pumps providing heating in new build houses in Germany \citep{vivid}. The Market Incentive Programme (MAP) for renewable heat has made investment grants available since 2008 for the installation of ground-source and air-to-water heat pumps, linked to minimum seasonal performance factors \citep{vivid, zimny2015polish}. Higher investment grants are available for ground source heat pumps (given their higher up front costs) in comparison to air source heat pumps. This policy is in line with our model which shows ground source heat pumps as economically favourable in Germany under higher temperature increases, see Fig. \ref{Fig: best_tech_bar}. While MAP also offers higher grants for heat pumps installed in existing homes compared to new builds, the share of heat pump sales to the retrofit sector has still been decreasing in recent years. This illustrates that increasing the portion of renewable heat in the existing housing stock continues to be a key priority \citep{vivid}. \\
 
\noindent Our results indicate that combined air-to-air heat pump/electric boiler systems are cost-optimal in Norway and Sweden under moderate temperature increases, see Fig. \ref{Fig: best_tech_bar}. In Norway, electricity price rises and investment subsidies for end users have stimulated the uptake of heat pumps \citep{vivid}. A subsidy programme for householders was introduced in 2003 which covered 20\% of the initial costs for installing air-to-air heat pumps, although this subsidy has since ended \citep{bjornstad2012diffusion, vivid}. In 2015 there were approximately 750.000 heat pumps installed in a third of all households in Norway, delivering 15 TWh of heat \citep{Patronen2017Nordic}. In Sweden, over half of heat pump sales in buildings in 2016 were for reversible air-to-air heat pumps, which can operate in conjunction with direct electric heating in existing homes, and are also used in commercial buildings \citep{european2014european, european2017european, Forsen2008Market}. Swedish building regulations requiring greater energy efficiency and lower heating demands in new buildings have contributed to the increasing dominance of air-to-air heat pumps compared to ground source heat pumps since 2005 \citep{zimny2015polish}. 

\subsection{Prospects for heat pumps: lessons from countries with natural gas heating }
\label{Sec: prospect}

\noindent In this paper, we present a cost-optimal distribution of decentralised heating technologies for Europe. The analysis does not reflect how country to country possibilities for shifting to a more optimal mix of heating technologies are constrained by path dependency and lock-in to existing heating infrastructure. Selection and replacement of heating technologies is dependent upon contingent factors such as the coverage and extent of gas networks, and availability of local natural resources for fuels and energy sources \citep{gross2019path, hanna2016best, vivid}.\\

\noindent Policy support aimed at achieving a transition from gas boilers to heat pumps is challenging in countries such as the Netherlands and the UK where gas heating is dominant. Both countries have made very limited progress in substituting gas heating for lower carbon alternatives. Our numerical findings show that ground source heat pumps are a definitively cost-optimal solution in the UK, see Fig. \ref{Fig: best_tech_bar}. However, the UK has a low uptake of heat pumps compared to many northern and western European countries. In part this is due to path dependent developments that led to the emergence of gas central heating as an affordable, convenient and familiar technology for UK household consumers, which provides high levels of thermal comfort \citep{gross2019path}. There is also a lack of continuous, coordinated policies on technical standards and promotion of heat pumps in the UK, leading to poor consumer awareness and low confidence in the technology \citep{hanna2016best, hanna2018microgeneration}. In order to help meet the 5th carbon budget \citep{committee2015fifth}, the UK's Committee on Climate Change has set a target for at least 2.5 million heat pumps to be installed in UK homes by 2030, compared to a total stock of approximately 160.000 in 2016 \citep{committee2018reducing}. \\

\noindent In contrast to the UK, the Netherlands has ambitious long-term policy targets to phase out the contribution of natural gas to its heat supply. Fig. \ref{Fig: best_tech_bar} suggests that in the Netherlands, heat pumps are cost-optimal under climate change scenarios. On other hand, approximately 85\% of dwellings in the Netherlands use natural gas for space heating \citep{european2017european}. The dominance of natural gas as a heating source was cemented after the discovery of extensive local gas reserves in the Groningen gas field in 1959 \citep{vivid}. Earthquakes in 2012 caused by gas extraction from the Groningen gas field led to public protests and the government therefore decided to reduce gas production \citep{honore2017dutch}. Depletion of the gas field is also expected, so that the Netherlands is likely to become a net importer of natural gas from 2030 to 2035 \citep{ECN2016}. Energy security concerns over the potential future need to import Russian gas, in addition to climate policy, have also motivated long-term national goals to seek low carbon alternatives to natural gas \citep{vivid}.\\
 
\noindent The 2050 Dutch Energy Agenda implies that most of the country's 6 million homes currently heated with natural gas will need to be disconnected from gas supply by 2050 \citep{european2017european, MEA2017}. This is based on a long-term target to reduce CO$_2$-emissions from the residential and commercial heating sector by 80-95\% by 2050 \citep{MEA2017}. The Dutch government has indicated that electrification could be central to plans to phase out natural gas heating, and have stated that energy efficiency, heat pumps and district heating would be three key elements to a low emissions building sector. Strategic decisions will need to be made about the future of the existing gas grid, in terms of whether it could be utilised as a carrier for alternatives to gas such as hydrogen and green gas, or whether it might need to be decommissioned and replaced \citep{vivid}. 

\subsection{Study limitations}
\label{Sec: stud_lim}

\noindent The analyses of this paper are subject to a number of limitations. The main limitation deals with the assumption of constant fuel and technology related prices across regions and time. As previously stated, this is a simple approximation which leads to an unrepresentative picture of the real world. However, this is necessary for performing an objective study of the influence of climate change, separate from the political and personal drivers, and thus answer the research questions of this work. \\

\noindent Secondly, the optimisation was limited to only consider cost constraints. Other important subjects to constrain are the different projections of RCP related CO$_2$-emissions. This limitation suggests that results belonging to otherwise low or negative CO$_2$-emission scenarios as, e.g., for the end-of-century RCP26 pathway, should be interpreted with caution. An in-depth analysis of the cost-optimal design decisions, including CO$_2$-emission constraints, would additionally require a coupling to the electricity sector. Such an extensive work require a study on its own. \\

\noindent Then, this study is limited to only consider decentralised heating technologies, such as small scale fossil fueled boilers or heat pumps. Large-scale centralised technologies, such as district heating, are also important to consider, but these are subject to different economic and political considerations, and thus cannot be thought of as a like-for-like replacement for existing heating systems in all countries of Europe. This study can be extended to include these by assigning heat transportation to densely populated areas as, e.g., cities or suburbs. \\

\noindent A fourth limitation is the relatively low number of climate models that are used to acquire the results. To the best of the authors' knowledge no other climate affected data exists with higher data granularity. If applicable, an identical analysis can be performed by including more climate data. \\

\noindent Our results do not consider the practicality of deploying technologies in different countries or locations given the relative coverage of gas or electricity networks, natural resource endowments, and the origins of path dependencies in incumbent heating systems. Therefore, in some countries or areas, certain heating technologies, even though being cost-optimal, might not be physically feasible to install because of the lack of distribution networks to meet the large demands for heating \citep{staffell2019role}. \\

\noindent Additionally, our analysis does not account for the costs of network infrastructures (also given their local availability) required to operate the heating technologies, i.e. gas and electricity networks. We recognise that network costs may vary significantly by location or depending on energy sources and vectors used to achieve heat decarbonisation \citep{strbac2018analysis}, and this could have a significant impact on our results in terms of overall system and end user technology costs. Also, location-specific barriers to technology uptake are not considered, such as lack of suitable building types or land availability for ground source heat pumps in urban areas. Together, these constraints could have a significant impact on results, and are recommended as areas for further study.

\section{Conclusions and policy implications}

\noindent This study set out to determine the cost-optimal design decision for decentralised heating in European homes. With fixed costs, we let the climate determine the most economical heating technologies over a timespan of 20 years in a historical period defined to span the years 1970-1990 and for three end-of-century time periods defined to span the years 2080-2100. Bias-adjusted high resolution temperature data from 9 combinations of 5 global climate models downscaled by 4 regional climate models under the EURO-CORDEX project, have been adapted to model the heat demand and supply side. Due to transparency issues on actual heat consumption data, we use the theory of heating degree-days as a proxy for the variations in the day-to-day heat demand over 20 years. Notwithstanding the impact of policies or consumer preferences, we have shown that the climate holds a vital role in the cost-optimal design of decentralised heating infrastructures in Europe. \\

\noindent We found that climate change introduces a decreasing demand for space heating. This demand is estimated to decrease by approximately 42\%, 24\% and 16\% in 2100 for the RCP8.5, RCP4.5 and RCP2.6 climate projections, respectively, compared to the corresponding value in 1970. Peak heating demand may have been overestimated in the UK for example \citep{watson2019decarbonising}, and can also be expected to reduce with the implementation of tighter building energy efficiency standards and a more thermally efficient European building stock by later in the century. In Europe, current EU and national heat policies and regulations do not appear to be future-proofed to account for potential long-term changes in heat demand, and how these changes might affect the optimal mix and economics of decentralised heating technologies. We recommend that similar to EU mandates for energy efficiency in buildings, it is important that heat policy frameworks take a flexible and technology neutral approach that allows for uncertain changes in heat demand due for example to climate change, energy efficiency policies, or increased electrification of heating. The calculation in the EU Renewable Energy Directive, which stipulates a minimum SPF for heat pumps to be considered a form of renewable energy \citep{gleeson2013meta}, may also need to be revised to account for the impact of lower end-use heating load on heat pump performance. \\

\noindent Climate change and cost-optimisation suggest a shift in the decentralised heating infrastructure from gas boilers to ground source heat pumps, for a majority of the European countries. This is driven by the two main factors. Firstly, climate change will increase ambient temperatures and thus improve the efficiency of heat pumps. Secondly, the diminishing need for space heating means the year-round demand for hot water becomes more important, increasing the utilisation of heating technologies through the year, thus benefiting heat pumps with their high upfront, but low ongoing, costs. For many countries this is in correspondence with aggressive policies on increasing penetrations of heat pumps. The far west and far south European countries are subjected to a high heat load factor, for which ground source heat pumps serve as economically favourable. As for the UK, there is a comparatively low uptake of heat pumps, where key barriers to further uptake are a lack of policy continuity and co-ordination and low consumer confidence and awareness. On the other hand, the Netherlands is developing ambitious long-term policies to increase the penetration of heat pumps. Portugal, Spain, Italy and Greece lack policy support for heat pumps, where the majority of sales are for reversible heat pumps which provide both heating and cooling. Aggressive policies for substantial deployment of heat pumps are evident in countries such as Sweden, Switzerland and Austria. For Switzerland, this is in correspondence with a higher penetration of ground source heat pumps at the end-of-century time frames in all RCPs. Climate conditions in Sweden and Austria lead to a marginal increase in the penetration of ground source heat pumps, while gas fired boilers take the largest technology share. High carbon taxes in Finland, Norway, Sweden and Denmark are used to pursue a decrease in the use of fossil fuel heating technologies. Only for Denmark do the climate conditions align with these policies. The continental climate across eastern Europe leads to a robust choice of gas boilers as the cost beneficial choice of heating in all RCPs. \\

\noindent Despite our findings suggesting a wider deployment of ground source heat pumps, the majority of heat pump sales in Europe over the last decade have been for air source heat pumps and in particular reversible air-to-air heat pumps which can provide both heating and cooling \citep{european2014european, european2017european}. Air source heat pumps tend to be cheaper and easier to install than ground source heat pumps, and have benefited from technical improvements, which have raised their efficiency and increased their suitability to perform effectively in a wider range of climatic conditions and building types \citep{european2017european}. \\

\noindent In general, ground source heat pumps are used to a greater extent in countries with colder climate zones, e.g., Nordic countries, where the heat source temperature needs to be more stable, although reversible air-to-air heat pumps still have the highest share of heat pump sales in these countries over the last ten years \citep{european2017european}. Reversible air-to-air heat pumps dominate sales in warmer, southern Mediterranean countries, where purchases of ground source heat pumps are minimal or absent. This overall pattern is by no means universal and the country-to-country distribution of ground source versus air source heat pumps is affected by various non-climate related factors affecting consumer uptake of heat pumps, including upfront costs, building regulations, availability of sufficient space for ground source heat pump components and thermal stores, and inconvenience caused by installation \citep{balcombe2014investigating, european2017european}.\\

\noindent We recognise that national heating policies aimed at decarbonisation should be designed with respect to a portfolio of technologies, including those not featured in this analysis, for example district heating or hydrogen. Our findings should therefore not be treated as a policy prescription, and moreover the analysis focuses on the impact of climate change on technology costs and excludes a calculation of carbon intensity. This carries the consequence that under higher temperature increases, gas heating is indicated as a more optimal technology in a country such as Sweden, which has promoted deep decarbonisation of its heating sector and where the heat pump market is mature. Meeting the 1.5 $\degree$C target under the Paris Agreement and net zero emissions reduction targets imply a significantly reduced or minimal role for natural gas in the energy mix. Gas heating could have a role as a bridging technology to low carbon heating for example through the use of hybrid heat pumps, whereby a heat pump operates in parallel with a gas boiler. \\

\noindent This study underlines the benefit of accelerating heat pump support policies in countries such as the UK and Netherlands which depend largely on natural gas heating. These countries can learn from the experience in northern and western Europe, where multiple policies have been combined over several decades to raise levels of heat pump deployment, including technical standards, promotion and quality assurance, up front investment subsidies, and carbon and energy taxes. While our study points to heat pumps being more cost-optimal in western Europe under projected climate change, heat pump markets in countries such as Austria, Switzerland, France and Germany are actually less mature than in northern Europe. Gas boilers are also identified as being more optimal in eastern Europe, where in some cases (such as Poland) there may be potential to displace more carbon intensive coal heating with natural gas heating. One of the most surprising outcomes of the study is that ground source heat pumps are identified as suitable solutions in southern Mediterranean countries. This points to a need for more coordinated policies in these countries to support heat pumps as a low carbon alternative for heating in buildings, given the current predominance of reversible heat pumps which are also used for air conditioning, and have emerged largely as a market development in the absence of specific policy assistance. 
\section{Nomenclature}

\begin{table}[H]
\begin{tabular}{|p{2cm}p{6cm}|}
\hline
\textbf{Subscripts} & \textbf{Explanatory text} \\ 
$x$ & Grid cell \\
$\Delta$ & Time period \\
$\theta$ & Technology \\ 
{} & {} \\ \hline
\textbf{Variables} & \textbf{Explanatory text} \\ 
$t$ & time \\
$L^\text{Total}$ & Total residential heat demand  \\
$L^\text{Space Heat}$ & Space heat demand \\
$L^\text{Hot Water}$ & Hot water demand  \\
$P^\text{Total}$ & Total output of heat \\
$P^\text{Space Heat}$ & Output of space heat \\
$P^\text{Hot Water}$ & Output of hot water  \\
$T_0$ & Heating threshold temperature \\
$T_x(t)$ & Gridded ambient air temperature\\
$T_{x,\text{ design}}$ & Gridded design temperature \\
$\mu_x$ & Gridded heat load factor \\
$\text{HDD}_x^{\text{Space}}(t)$ & Gridded heating degree-days as a proxy for space heating \\
DHW & Heating degree-days as a proxy for hot water demand \\
$\text{HDD}_{x,\Delta}$ & Gridded heating degree-days as a proxy for the combined space heating and hot water over a time period \\
$\text{X}^{\text{CAP}}_\theta$ & Fixed capacity expense \\
$\text{X}^{\text{FM}}_\theta$ & Fixed yearly maintenance expense and auxiliary electricity use \\
$\text{X}^{\text{OP}}_{x,\theta}$ & Gridded operational expense  \\
$\text{X}^{\text{TOT}}_{x,\theta}$ & Gridded yearly accumulated expense \\
$\text{X}^{\text{I}}_\theta$ & Fixed installation expense \\
$\text{X}^{\text{Fuel}}_\theta$ & Fixed fuel price \\
$\eta_{x,\theta}$ & Technology efficiency \\
$\kappa_\theta$ & Installed technology capacity \\
$\text{COP}_{x,\theta}$ & Gridded coefficient of performance \\
$T_{\text{source}}$ & Temperature of hot reservoir \\
$T_{\text{sink}}$ & Temperature to be met by technology \\ \hline
\end{tabular}
\end{table}
\section*{Acknowledgement}
\noindent The authors wish to express their gratitude to O. B. Christensen and F. Boberg from the Danish Meteorological Institute for their contribution to understanding the climate outcomes and for supplying out data from the regional climate model HIRHAM5. G. Nikulin from the Swedish Meteorological and Hydrological Institute is thanked for supplying out data from the regional climate model RCA4. E. van Meijgaard from the Royal Netherlands Meteorological Institute is thanked for supplying out data from the regional climate model RACMO22E. P. Lenzen from the German Climate Computing Center is thanked for supplying out data from the regional climate model CCLM4. Parts of the EURO-CORDEX climate data have been acquired via ESGF data nodes. We thank Fabian Levihn from Stockholm Exergi for providing district heating consumption data. Thanks to Aarhus University Research Foundation for funding S. Kozarcanin with funding number AUFF-E-2015-FLS-7-26. I. Staffell acknowledges the Engineering and Physical Sciences Research Council for funding the IDLES project (EP/R045518/1). G. B. Andresen was funded by the RE-INVEST project, which is supported by the Innovation Fund Denmark under grant number 6154-00022B. R. Hanna and R. Gross were funded by the UK Energy Research Centre - Phase 3 (UK Research and Innovation Energy Programme, grant reference EP/L024756/1), the Committee on Climate Change and the Department for Business, Energy \& Industrial Strategy. 

\section*{Author Contributions} 
\noindent IS, GBA and SK designed the study. IS and GBA furthermore supervised the entire study. SK administrated the project, performed the scientific investigation, wrote the majority of the paper apart from Section 4, performed data validation and visualisation. SK wrote the supplementary information. RH produced Section 4 on the political aspects in collaboration with RG and drawing on research undertaken by RH and RG for UKERC. 

\section*{Correspondence} 
\noindent Correspondence should be addressed to S. Kozarcanin (email: sko@eng.au.dk) or G. B. Andresen (email: gba@eng.au.dk).

\section*{Competing financial interests}
\noindent The authors declare no competing financial interests.

\section*{Bibliography}
\bibliographystyle{elsarticle-harv} 
\bibliography{bib}

\clearpage



\section*{Supplementary material}

# Impact of climate change on the cost-optimal mix of decentralised heat pump and gas boiler technologies in Europe

S. Kozarcanin[a,b,∗], R. Hanna[c], I. Staffell[b,c], R. Gross[c], G. B. Andresen[a]

[a]*Department of Engineering, Aarhus University, Inge Lehmanns Gade 10, 8000 Aarhus, Denmark*
[b]*Imperial College Centre for Environmental Policy, 16–18 Prince's Gardens, South Kensington, London SW7 1NE, UK*
[c]*Imperial College Centre for Energy Policy and Technology, 16–18 Prince's Gardens, South Kensington, London SW7 1NE, UK*

**This document is the supplementary information (SI) for the paper: Impact of climate change on the cost-optimal mix of decentralised heat pump and gas boiler technologies in Europe.**

The material of the SI is intended to give a broader perspective on the methods that are applied to conduct the technical analysis of this paper. It is not intended for the regular reader, but only for other researchers, with special interests. Section 1.1 provides a general overview of the heating degree-days and how it used as a proxy to model the day-to-day fluctuations in the heat demand. Section 1.2 presents our approach to model the supply side of the residential heating sector. In section 1.3, we describe our approach to model the Coefficient Of Performance (COP) of heat pumps. Section 1.4 presents the climate data that is implemented into this study. In Section 2.1 we extend our discussion on the *unperturbed* pricing scheme. Finally, in Section 2.2 we discuss our results on the impact of climate change on the heat pump coefficient of performance.

**Contents**

[∗]Corresponding author
 *Email addresses:* sko@eng.au.dk (S. Kozarcanin), r.hanna@imperial.ac.uk (R. Hanna), i.staffell@imperial.ac.uk (I. Staffell), robert.gross@imperial.ac.uk (R. Gross), gba@eng.au.dk (G. B. Andresen)

# 1. Extended Methodology

## 1.1. Heat load factors

The heat load factor, denoted as $\mu$, is defined as the unitless ratio of the residential heat demand, $L^{\text{Total}}$, to the maximum possible output of heat, $P^{\text{Max}}$, over a given period of time, $\Delta$, as:

$$\mu = \frac{L^{\text{Total}}}{P^{\text{Max}} \cdot \Delta} \quad (1)$$

where the total residential heat demand, $L^{\text{Total}}$, is the sum of the individual space heating and hot water components, $L^{\text{Space Heat}}$ and $L^{\text{Hot Water}}$, respectively:

$$L^{\text{Total}} = L^{\text{Space Heat}} + L^{\text{Hot Water}}$$

The decentralised nature of heating means that data on consumption is not readily available and therefore not applicable for further research. Known to the literature, the theory of heating degree-days is most frequently used as a best proxy for the variations in the day-to-day heat demand as, e.g., in Berger and Worlitschek (2018) and Christenson et al. (2006). In this study, we assume a direct proportionality between the total residential space heat demand, $L^{\text{Total}}$, and the heating degree-days, $\text{HDD}^{\text{Space Heat}}$, as:

$$L^{\text{Total}} = L^{\text{Space Heat}} + L^{\text{Hot Water}} \quad (2)$$
$$= \alpha \cdot \text{HDD}^{\text{Space Heat}} + L^{\text{Hot Water}}$$

$\alpha$ is a constant of proportionality in units of energy per heating degree-day. Inspired by Kozarcanin et al. (2019), the accumulated heating degree-days, $\text{HDD}^{\text{SpaceHeat}}_{\Delta,\text{x}}$, for a single grid location, x, over a period of time, $\Delta$, is given as:

$$\text{HDD}^{\text{Space Heat}}_{\Delta,\text{x}} = \int_{\Delta} (T_0 - T_{\text{x}}(t))_+ \, dt \quad (3)$$

$(T_0 - T_{\text{x}}(t))_+$ defines a positive value or otherwise zero (Thom, 1954). To elaborate, if $T_0 > T_{\text{x}}(t)$, the output of $(T_0 - T_{\text{x}}(t))_+$ will add to the heating degree-days. On the other hand, if $T_0 \leq T_{\text{x}}(t)$, $(T_0 - T_{\text{x}}(t))_+$ will be put to zero. $(T_0 - T_{\text{x}}(t))_+$ is defined as:

$$(T_0 - T_{\text{x}}(t))_+ = \begin{cases} T_0 - T_{\text{x}}(t) & \text{if } T_0 > T_{\text{x}}(t) \\ 0 & \text{if } T_0 \leq T_{\text{x}}(t) \end{cases}$$

The base temperature, $T_0$, is defined as the outside temperature below which a building is assumed to need heating. For simplicity, the base temperature is assumed to be 17 °C although strong evidence suggests that this value vary according to region and study (Kozarcanin et al., 2019). $T_{\text{x}}(t)$ defines the gridded time series of the ambient air temperature.

The maximum output of heat, $P^{\text{Max}}$, is in a similar way to the total heat consumption defined by a maximum output of space heat, $P^{\text{Space Heat}}$, and hot water, $P^{\text{Hot Water}}$, as:

$$P^{\text{Max}} = P^{\text{Space Heat}} + P^{\text{Hot Water}} \quad (4)$$
$$= \frac{\alpha \left(T_0 - T_{\text{x, design}}\right) \cdot 1\text{day}}{\Delta} + \frac{L^{\text{Hot Water}}}{\Delta}$$

The maximum output of hot water, $P^{\text{Hot Water}}$, equals the consumption of hot water, $L^{\text{Hot Water}}$, normalised to $\Delta$. $P^{\text{Space Heat}}$ is defined in a similar way to $L^{\text{Space Heat}}$. $T_{\text{x, design}}$ is the system design temperature and calculated as a 0.05% quantile of the gridded daily ambient temperature, $T_{\text{x}}(t)$, as:

$$\int_0^{T_{\text{x, design}}} p_{\text{x}}(T_{\text{x}}) \, dT_{\text{x}} = 0.0005$$

A small quantile is used to ensure an operation time in conditions above the design temperature in 95.95% of the time. A 0.05% quantile corresponds to approximately 5 hours during a year. A 100% quantile would otherwise overestimate the technology capacity and increase the capital investments.

Finally, replacing $L^{\text{Total}}$ and $P^{\text{Max}}$ in Eq. 1 by the expressions given in Eq. 2 and 4 leads to:

$$\mu_{\text{x}} = \frac{\alpha \cdot \text{HDD}^{\text{Space Heat}}_{\Delta,\text{x}} + L^{\text{Hot Water}}}{\left(\frac{\alpha(T_0 - T_{\text{x, design}}) \cdot 1\text{day}}{\Delta} + \frac{L^{\text{Hot Water}}}{\Delta}\right)\Delta}$$

This can be simplified by removing the $\Delta$ in the denominator and dividing both the numerator and denominator by $\alpha$ as:

$$\mu_{\text{x}} = \frac{\text{HDD}^{\text{Space Heat}}_{\Delta,\text{x}} + \frac{L^{\text{Hot Water}}}{\alpha}}{(T_0 - T_{\text{x, design}}) \cdot 1\text{day} + \frac{L^{\text{Hot Water}}}{\alpha}} \tag{5}$$

Much like the energy demand for space heating, measurements of hot water consumption, $L^{\text{Hot Water}}$, are not available on scales that match the needs of this study. Therefore in Eq. 5, $\frac{L^{\text{Hot Water}}}{\alpha}$, which is given in units of heating degree-days, is used as a best proxy for hot water consumption. To provide a best approximation of $\frac{L^{\text{Hot Water}}}{\alpha}$, we use measured hot water and space heat consumption data from Stockholm (Levihn, 2018). The general logic is to estimate the ratio of hot water to space heat for Stockholm and use this value to estimate $\frac{L^{\text{Hot Water}}}{\alpha}$. Based on the measured consumption data, the ratio of hot water to space heat is 26%, i.e.:

$$L^{\text{Hot Water}}_{\text{Stockholm}} = 0.26 \cdot L^{\text{Space Heat}}_{\text{Stockholm}}$$
$$L^{\text{Hot Water}}_{\text{Stockholm}} = 0.26 \cdot \alpha \cdot \text{HDD}^{\text{Space Heat}}_{\Delta,\text{Stockholm}}$$
$$\frac{L^{\text{Hot Water}}_{\text{Stockholm}}}{\alpha} = 0.26 \cdot \text{HDD}^{\text{Space Heat}}_{\Delta,\text{Stockholm}}$$

Depending on the time extent, $\frac{L^{\text{Hot Water}}_{\text{Stockholm}}}{\alpha}$ can be estimated simply by evaluating Eq. 3. If data from other locations are known, these can in an identical way be used to estimate the heating degree-day proxy of the hot water consumption. Finally, we assume a constant hot water consumption across space and time, which means that $\frac{L^{\text{Hot Water}}}{\alpha}$ is fixed to the value for Stockholm for each grid location, x.

*1.2. Techno-economic standpoint of heat generation*

The hourly accumulated cost, $X^{\text{TOT}}_{\text{x},\theta}$, for a technology, $\theta$, and grid location, x, depends linearly on the heat load factor, $\mu_{\text{x}}$, as:

$$\text{X}^{\text{TOT}}_{\text{x},\theta} = \text{X}^{\text{CAP}}_{\theta} + \mu_{\text{x}} \cdot \text{X}^{\text{OP}}_{\text{x},\theta}$$

The capital expense, $\text{X}^{\text{CAP}}_{\theta}$, is assumed to be proportional to the installed capacity, $\kappa_\theta$, as shown in Eq. 6. The per MW equipment, installation and maintenance expenses are denoted as $\text{x}^{\kappa}_\theta$, $\text{x}^{\text{I}}_\theta$ and $\text{x}^{\text{FM}}_\theta$, respectively. The capital cost is annuitised by the technology life time and a discount rate of 4%.

$$\text{X}^{\text{CAP}}_\theta = \left(\text{x}^{\text{FM}}_\theta + \text{x}^{\text{I}}_\theta + \text{x}^{\kappa}_\theta\right) \cdot \kappa_\theta \tag{6}$$

The marginal expense, $X_{x,\theta}^{OP}$, is proportional to the installed capacity, $\kappa_\theta$ as well as the ratio between the fuel price, $x_\theta^{Fuel}$, and efficiency, $\text{eff}_{x,\theta}$:

$$X_{x,\theta}^{OP} = \frac{x_\theta^{Fuel}}{\text{eff}_{x,\theta}} \cdot \kappa_\theta$$

The efficiency takes a constant value, $\eta_\theta$, for technologies different from heat pumps, as:

$$\text{eff}_{x,\theta} = \begin{cases} \text{COP}_{x,\theta}(t) & \text{if } \theta \text{ defines a heat pump} \\ \eta_\theta & \text{otherwise} \end{cases}$$

where $t$ defines the time. All prices and technology properties are given in Tab. 1 in the main paper.

### 1.3. Coefficient of performance (COP)

Heat pumps are implemented with a Coefficient Of Performance (COP), which defines the ratio of heat output to the amount of electricity input. COP is strongly temperature dependent, thus, long-term average values are not meaningful. An empirical relationship between the COP and the temperature difference between the heat source and heat sink, $\Delta T = T_{sink} - T_{source, x}(t)$, is presented in Eq. 7 for air and ground based heat pumps, respectively. These are based upon the derivations in Staffell et al. (2012), updated to include new data from (NTB Buchs, 2019). Furthermore, the COP for air source heat pumps was separated based on whether or not defrosting is required. Defrosting is required when outdoor temperatures fall below 5°C, lowering the COP by around 4%. For air and ground source heat pumps, $T_{source, x}$ represents the gridded air and soil temperatures, respectively. The ground temperature is estimated as an average of air temperatures over a 20 year time period. As discussed in the main paper, this corresponds to temperatures at a depth of approximately 50 meters below ground, depending on soil type and geographical location (GIAU/GEUS, 2014). $T_{sink}$ is assumed to be 30°C for air to air heat pumps and 55°C for large area hot water heating Staffell et al. (2012).

$$\text{COP}_{\text{Air driven, x}}(t) = \begin{cases} 0.0012\Delta T^2 - 0.1702\Delta T + 7.855 & \text{if } T_{\text{air}} \leq 5°C \\ 0.0019\Delta T^2 - 0.2258\Delta T + 9.073 & \text{if } T_{\text{air}} > 5°C \end{cases}$$
$$\text{COP}_{\text{Ground driven, x}}(t) = 0.0019\Delta T^2 - 0.2544\Delta T + 11.008 \quad (7)$$

During winter periods, when the need for heat is high, COP takes lower values. Oppositely, during summer periods the need for heat is small but the COP increases considerably. Consequently, the yearly averaged COP for air-to-air and air-to-water heat pumps, $\text{COP}_{\text{ASHP, x}}$, is weighted by the heating degree-days for space heating, $\text{HDD}_x^{\text{Space Heat}}(t)$, as shown in Eq. 8. The hot water component makes no difference to the weighting due to the assumption of constant consumption throughout the year. A similar weighting is not necessary for the ground-to-water heat pumps as temperatures at a depth of 50 meters below ground are seasonally independent (GIAU/GEUS, 2014).

$$\text{COP}_{\text{Air driven, x}} = \frac{1}{\text{HDD}_{x,\Delta}^{\text{Space Heat}}} \sum_{t \in \Delta} \text{HDD}_x^{\text{Space Heat}}(t) \cdot \text{COP}_{\text{Air driven, x}}(t) \quad (8)$$

### 1.4. Climate model temperature data

The combinations of regional and global climate models are shown in Tab. 1.

Table 1: Overview of the CMIP5 climate models implemented into this study along with the available climate projections, RCP26, RCP45 and RCP85.

| GCM | RCMs | Projections | | | |
|---|---|---|---|---|---|
| | | Historical | RCP26 | RCP45 | RCP85 |
| ICHEC-EC-EARTH | HIRHAM5 | x | x | x | x |
| | RACMO22E | x | | x | x |
| | RCA4 | x | x | x | x |
| MOHC-HadGEM2-ES | RACMO22E | x | x | x | x |
| | RCA4 | x | x | x | x |
| MPI-ESM-LR | CCLM-8-17 | x | | x | x |
| | RCA4 | x | x | x | x |
| IPSL-CM5A-MR | RCA4 | x | | x | x |
| CERFACS-CNRM-CM5 | RCA4 | x | | x | x |

Key figures of the three projections, RCP2.6 (Vuuren et al., 2011), RCP4.5 (Thomson et al., 2011) and RCP8.5 (Riahi et al., 2011) are shown in Fig. 1. RCP2.6 is reflected by the Paris agreement, to keep the global temperature rise well below 2°C above pre-industrial levels at the end of the 21st Century (Agreement, Paris, 2015), as seen in Fig. 1 panel c. This can be realised through the green peak-and-decline pathways of $CO_2$-emissions and concentrations, as shown in panels a and b, respectively. Oppositely, RCP8.5 represents a future with an increase of $CO_2$-emissions as of today. The result of the consequent $CO_2$-concentrations is a European average temperature increase of up to 5°C. RCP4.5 defines an intermediate scenario with stringent climate policies such as economic penalties for $CO_2$-emissions.

The ability of the Global Climate Models (GCM) to accurately model the near surface air temperatures is receiving increasing attention. A recent study by Cattiaux et al. (2013) on the European domain shows negatively biased winter temperatures in the North for 33 CMIP5 GCMs compared to ground observations from ECAD (Van Den Besselaar et al., 2015). Positively biased summer temperatures are observed in the East and Central Europe. The GCM ensemble mean bias is approximately −1°C ±9°C during winter months and 0.5°C ±6°C during summer months. Similar trends are found for the Northern Eurasia where the winter and summer periods show the largest biases (Miao et al., 2014). Small improvements have been made since CMIP3 GCMs (Meehl et al., 2007). To address these issues, a bias adjustment approach is adapted from Kozarcanin et al. (2019) and used to bias adjust the temperature profiles.

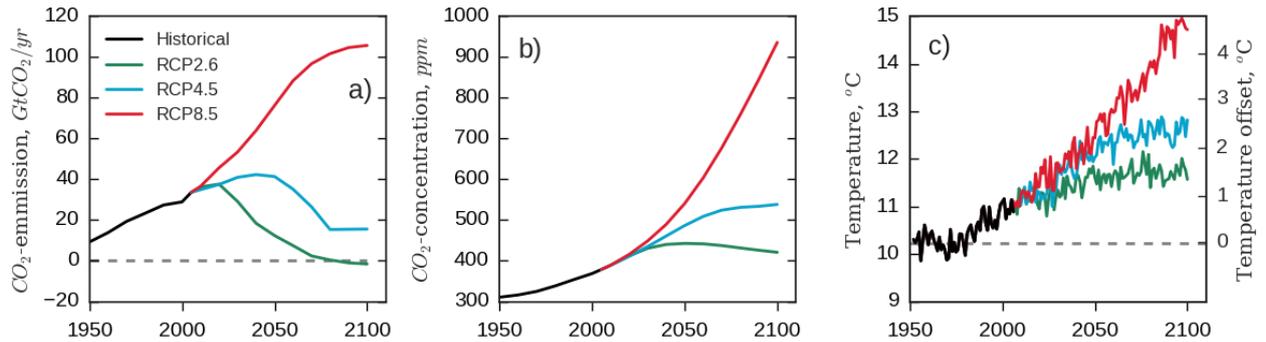

Figure 1: Key metrics of the IPCC climate projections. Panel a) presents the projected $CO_2$-emissions in Gt$CO_2$/yr. Panel b) presents the consequent concentration of $CO_2$ in the atmosphere in ppm. Panel c) presents the final temperature increases for Europe computed by HIRHAM5-ICHEC-EC-EARTH model. The right axis shows the temperature increase with respect to the 1950-1970 averaged European temperature.

## 2. Supplementary results

*2.1. Extended discussions on the original pricing scheme*

In this section, we further discuss the unperturbed pricing scheme and the single technology dominance across Europe.

Fig. 2 shows the screening curves for all technologies that are included in this study. Since the heat pump coefficients of performance fluctuate according to the ambient temperatures, as demonstrated by Eq. 7 and Fig. 4-6, heat pumps are subject to a range of screening curves. Fig. 4-6 are discussed in detail in Section 2.2. Taking the air-to-water heat pump as an example, it is seen from Fig. 4 that the coefficients of performance fluctuate between 2.0 and 3.0. As a consequence, air-to-water heat pumps are subject to an upper and lower screening curve that define the cost region, as illustrated in Fig. 2. Similar arguments can be made for the soil-to-water and air-to-air heat pumps. On the other hand, uniform efficiencies of biomass, oil and gas boilers result in a single screening curve for these technologies. The black shaded region defines the range of heat load factors across Europe. A spatial distribution of the heat load factors is shown in Fig. 3 and discussed in detail in the next paragraph. From Fig. 2 it is clear that within the range of the heat load factors, only gas boilers qualify as cost-optimal. This singularity lead to the *balanced* pricing scheme, which is used to enforce a more diverse technology distribution. This is needed in order to illustrate the potential impact of climate change on the heat generating technologies.

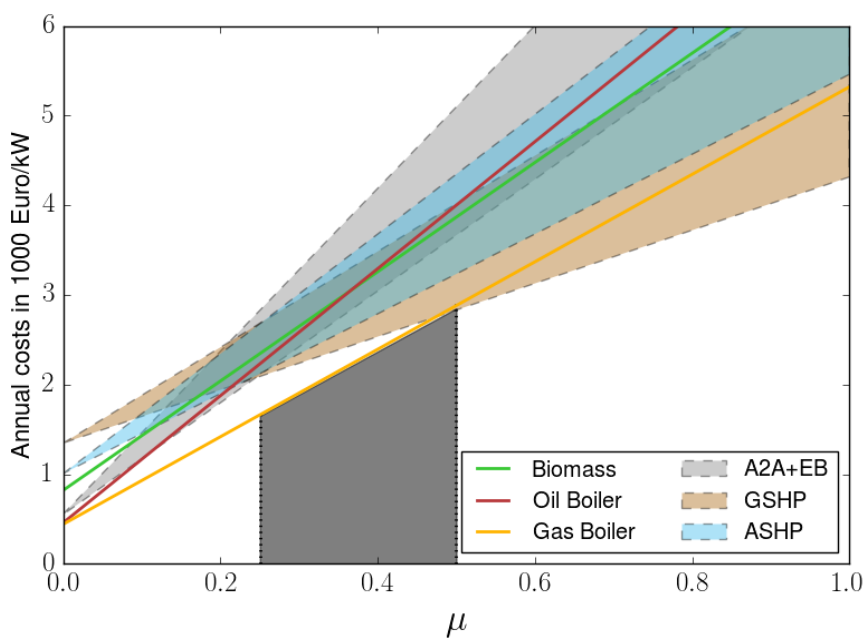

Figure 2: Screening curves showing the annual accumulated costs of heating in 1000 Euro/kW as a function of the heat load factor, $\mu$. Technology prices and technology properties are taken from Tab. 1 in the main article. The upper and lower screening curves for air-to-water heat pumps (ASHP) are defined by coefficients of performance equal to 2.0 and 3.0, respectively. The similar for the ground-to-water heat pumps (GSHP) are 2.5 and 4.5. For the hybrid system of air-to-air heat pumps and electricity driven boilers the combined efficiency are 3.0 and 5.0. The region shaded by a black color is constrained between 0.25 and 0.50 and defines the range of the heat load factors across Europe.

Fig. 3 shows the heat load factors across Europe for the historical period and for the end-of-century periods for each climate projection. Focusing initially on the historical frame, it is clear that the cold oceanic climate increases the heat load factors significantly across the British Isles. The similar is evident for Scandinavia. The Iberian Peninsula is as well dominated by high heat load factors. This is a result of a warm Mediterranean climate, which may seem contradicting. However, the increased temperatures across these regions

reduce naturally the need for space heating. As a consequence, the constant hot water consumption takes up a significant share of the total heat demand and in turn increasing the heat load factors.

A detailed investigation of Fig. 3 reveals that it is difficult to assign any trend to the heat load factors as a function of the degree of climate change. The change in heat load factors results from a combined effect of changes in the heating degree-days and changes in the design temperatures, as seen from Eq. 5. The Balkan countries possess almost the same heat load factors, while other parts of Europe are significantly affected by changes in the ambient temperatures. Modest temperature increases at the end-century of the RCP2.6 climate projection lead naturally to modest changes in the heat load factors. The intermediate temperature increase at the end-century of the RCP4.5 climate projection decreases the heat load factors to some extent. This change is mainly observed across the British Isles and Scandinavia. The extreme temperature increase at the end-century of the RCP8.5 climate projection leads to a significant decrease in the heat load factors in some parts of Europe as, e.g., across the British Isles, while other parts as, e.g., the Iberian Peninsula and East Europe stay almost unaffected.

*2.2. Extended discussions on the heat pump coefficients of performance*

Fig. 4 - 6 show the spatial distributions of the coefficients of performance for the three types of heat pumps in this study. Focusing initially on the air-to-water coefficients of performance in the historical time frame, it is clear that Scandinavia holds the lowest values. Higher ambient temperatures across the Mediterranean result in the highest values. In general, the values range between 2.0 and 2.7. The modest temperature increase at the end-century of the RCP2.6 climate projection does not lead to significant changes in the coefficient of performance. On the other hand, the end-century of the RCP4.5 and RCP8.5 lead to a significant increase in the coefficients of performance. In RCP8.5, the coefficients of performance increase by up to 2.5 in Scandinavia and up to 3.0 in the southern Iberian Peninsula. These changes contribute significantly to the increased distribution of heat pumps across Europe at the end-century of each climate projection. Similar arguments can be made for the ground-to-water and air-to-air heat pumps, as seen in Fig. 5 and 6, respectively.

Comparing the coefficients of performance from the different heat pumps, it is evident that the values for ground-to-water heat pumps are significantly higher compared to air-to-water heat pumps. This is mainly reasoned by the stable ground temperatures, which provide a high coefficient of performance independent of the yearly seasons. The low sink temperature of the air-to-air heat pumps results in the highest coefficient of performance.

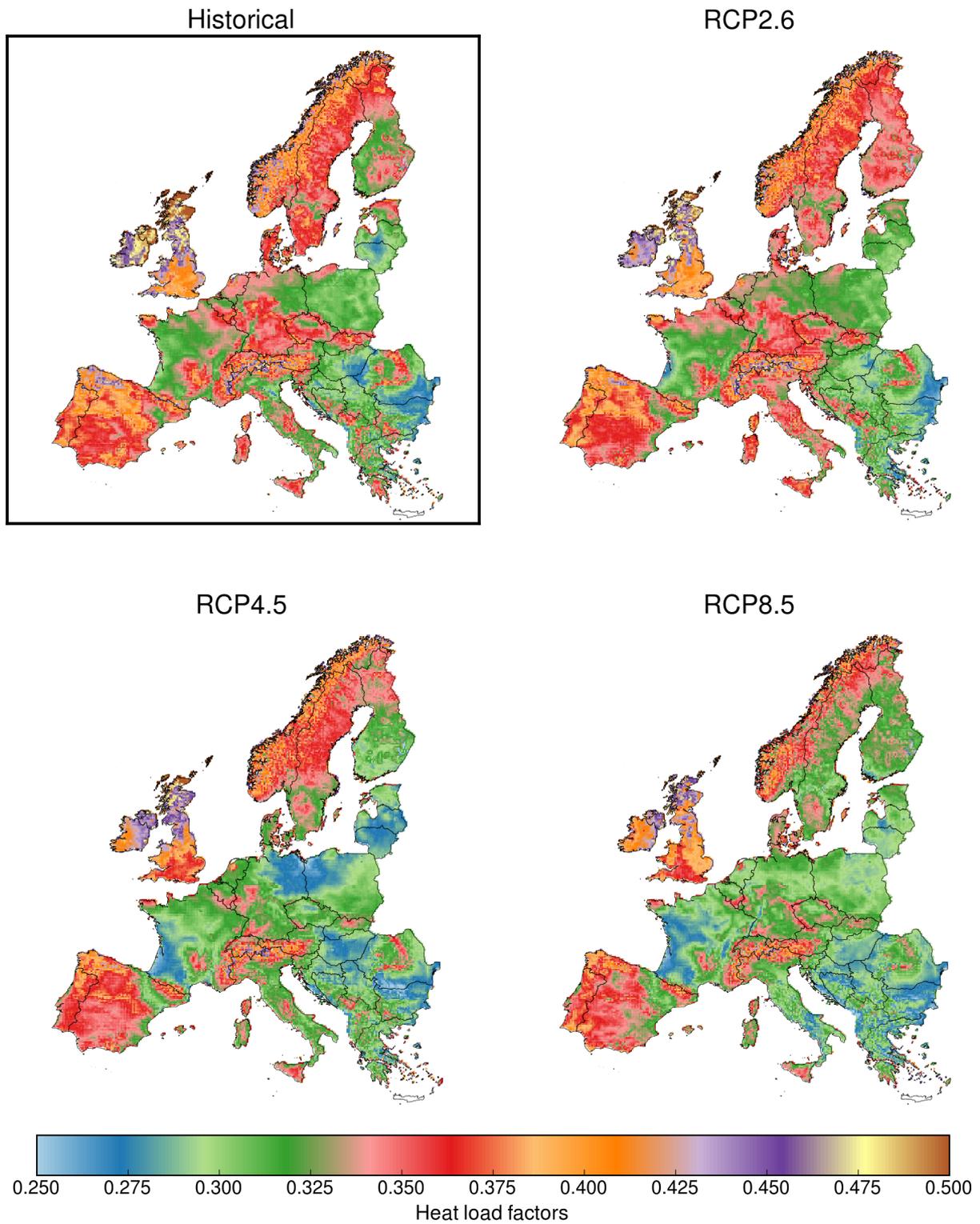

Figure 3: Spatial distributions of the heat load factors. The historical period is defined to span the years 1970-1990. RCP2.6, RCP4.5 and RCP8.5 spans a climatic period from 2080-2100. The figures are based on the ICHEC-EC-EARTH HIRHAM5 climate model.

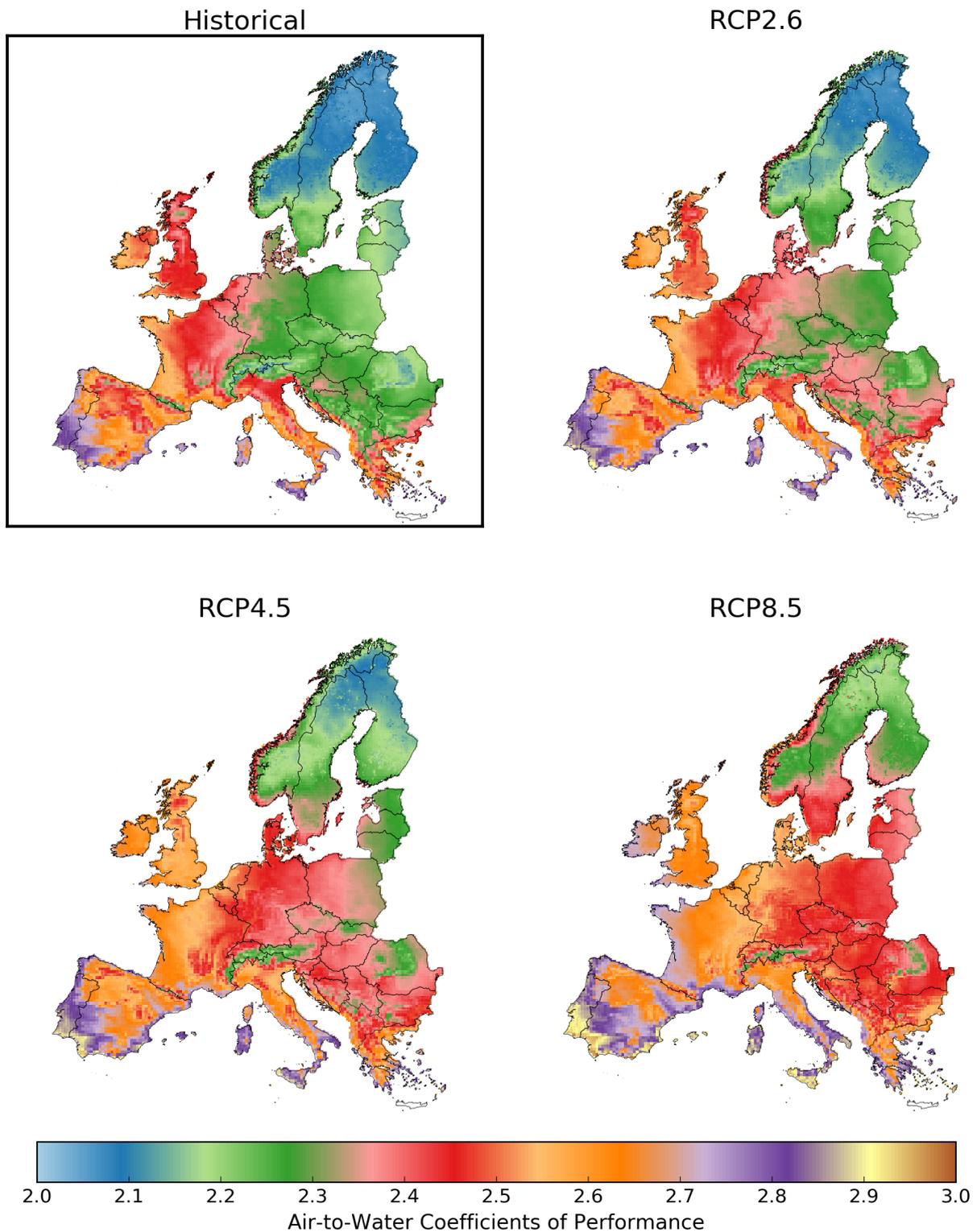

Figure 4: Spatial distributions of the coefficients of performance for air-to-water heat pumps with a sink temperature of 55 °C. These are weighted according the annual space heat demand as demonstrated in Eq. 8. The historical period is defined to span the years 1970-1990. RCP2.6, RCP4.5 and RCP8.5 spans a climatic period from 2080-2100. The figures are based on the ICHEC-EC-EARTH HIRHAM5 climate model.

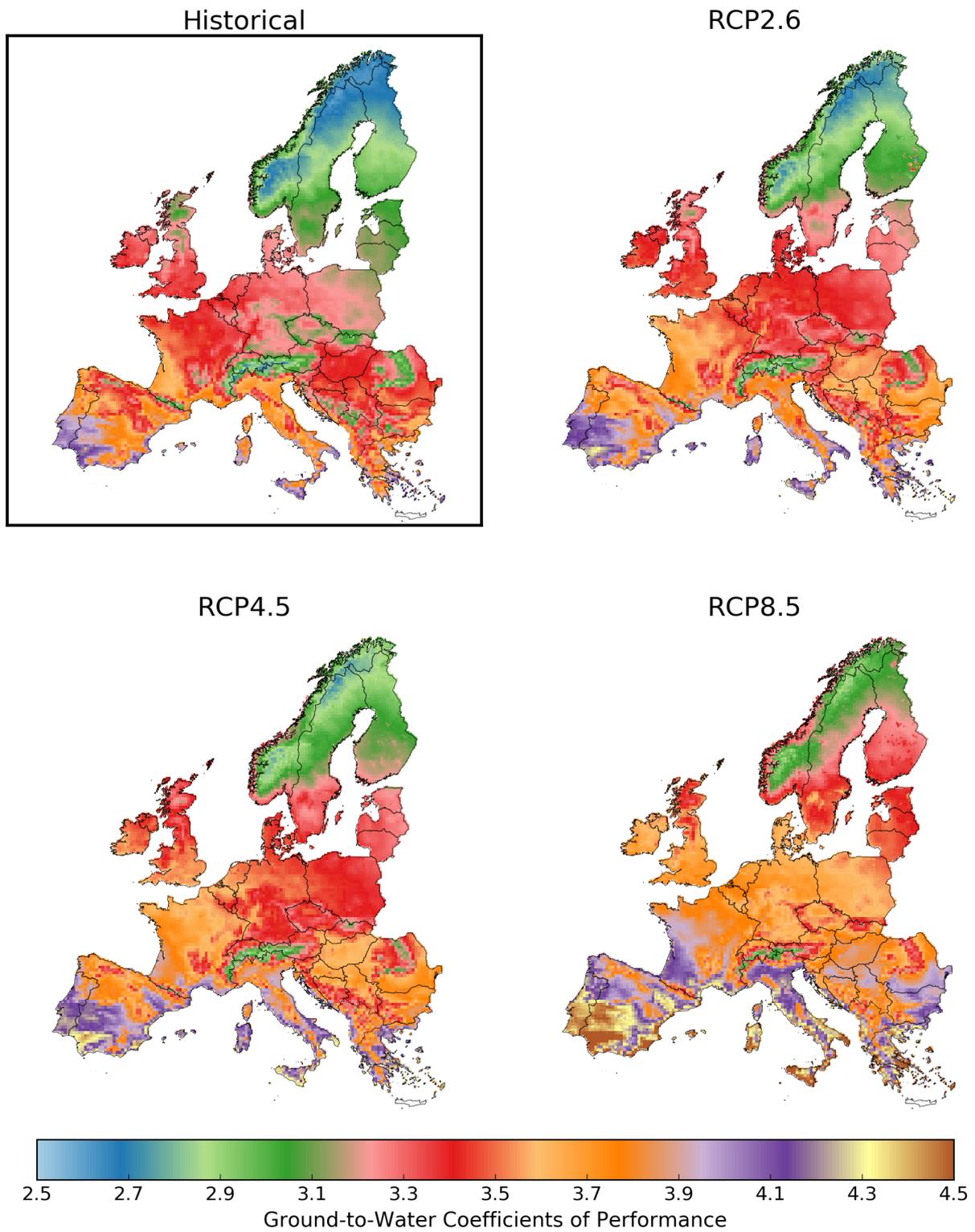

Figure 5: Spatial distributions of the coefficients of performance for ground-to-water heat pumps with a sink temperature of 55 °C. The historical period is defined to span the years 1970-1990. RCP2.6, RCP4.5 and RCP8.5 spans a climatic period from 2080-2100. The figures are based on the ICHEC-EC-EARTH HIRHAM5 climate model.

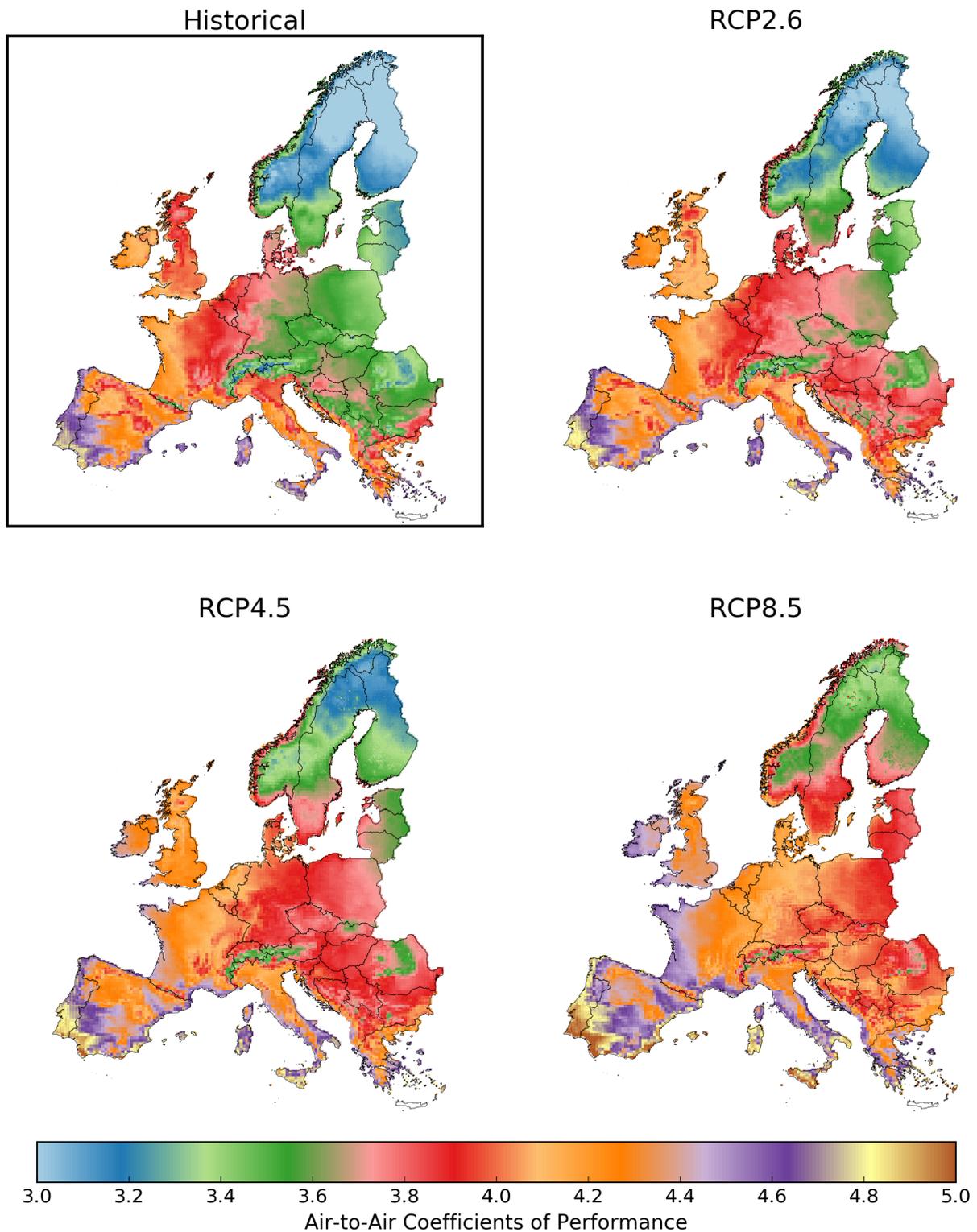

Figure 6: Spatial distributions of the coefficients of performance for air-to-air heat pumps with a sink temperature of 30 °C. These are weighted according the annual space heat demand as demonstrated in Eq. 8. The historical period is defined to span the years 1970-1990. RCP2.6, RCP4.5 and RCP8.5 spans a climatic period from 2080-2100. The figures are based on the ICHEC-EC-EARTH HIRHAM5 climate model.



\end{document}